\documentclass[a4paper]{article}

\usepackage{INTERSPEECH2018}
\usepackage{amsmath,graphicx}
\usepackage{amssymb,bm}
\usepackage{textcomp}
\usepackage{hyperref}
\usepackage[ruled,vlined]{algorithm2e}
\usepackage{color}
\usepackage{cite}

\title{Frequency domain variants of velvet noise and\\ their application to speech processing and synthesis: with appendices}
\name{Hideki Kawahara$^1$, Ken-Ichi Sakakibara$^2$, Masanori morise$^3$,\\ Hideki Banno$^4$, Tomoki Toda$^5$, Toshio Irino$^1$}
\address{
  $^1$Wakayama University, Japan\\
  $^2$Health Science University of Hokkaido, Japan\\
  $^3$University of Yamanashi, Japan\\
  $^4$Meijo University, Japan\\
  $^5$Nagoya University, Japan}
\email{kawahara@sys.wakayama-u.ac.jp, kis@hoku-iryo-u.ac.jp, mmorise@yamanashi.ac.jp, banno@meijo-u.ac.jp, tomoki@icts.nagoya-u.ac.jp}

\begin{document}

\maketitle
\begin{abstract}
We propose a new excitation source signal for VOCODERs and an all-pass impulse response for post-processing of synthetic sounds and
pre-processing of natural sounds for data-augmentation.
The proposed signals are variants of velvet noise, which is a sparse discrete signal consisting of a few non-zero (1 or -1) elements and
sounds smoother than Gaussian white noise.
One of the proposed variants, FVN (Frequency domain Velvet Noise)  applies the procedure to generate a velvet noise on the cyclic frequency domain
of DFT (Discrete Fourier Transform).
Then, by smoothing the generated signal to design the phase of an all-pass filter followed by inverse Fourier transform yields the
proposed FVN.
Temporally variable frequency weighted mixing of FVN generated by frozen and shuffled random number provides
a unified excitation signal which can span from random noise to a repetitive pulse train.
The other variant, which is an all-pass impulse response, significantly reduces ``buzzy'' impression of VOCODER output
by filtering.
Finally, we will discuss applications of the proposed signal for watermarking and psychoacoustic research.
\end{abstract}

\section{Introduction}
The Velvet noise is a sparse discrete signal which consists of fewer than 20\% of non-zero (1 or -1) elements.
The name ``velvet'' represents its perceptual impression.
It sounds smoother than Gaussian white noise\cite{jarvelainen2007reverberation,valimaki2013ieetr}.
We found that the frequency domain variants of velvet noise provide useful candidates for the excitation source signals of
synthetic speech and singing\cite{kawahara2018sigmus118}.
They can replace excitation source signal models\cite{kawahara1999spcom,Kawahara2001maveba,kawahara2010simplification,degottex2017ieeetr} for
 VOCODERs\cite{kawahara1999spcom,kawahara2008icassp,morise2016world}
and provide a unified design procedure of mixed-mode excitation signals.
The proposed frequency variant of the velvet noise is also an impulse response of an all-pass filter\cite{oppenheimerBook}.
The all-pass filter use of the frequency domain variant provides an effective and easy way for reducing ``buzzy'' impression of VOCODER speech sounds.
The impulse response of the variant is also a TSP (Time Stretched Pulse) and applicable to information hiding for tampering detection.
This article introduces the frequency domain variants of velvet noise
and discusses their use in speech signal processing including singing and speech synthesis.

\section{Background and related work}
How to analyze and generate the random component for synthetic voice has been 
a difficult problem\cite{yegnanarayana1998ieeetr,Kawahara2001maveba,Malyska2008,degottex2017ieeetr}.
In addition to this difficulty in analysis and synthesis, auditory perception introduces another difficulty.
It is the significant variation of the masking level of a burst sounds within one pitch period\cite{Skoglund2000ieeetrans}.
The reference suggests that two synthetic speech sounds having 20~dB SNR difference provide perceptually equivalent SNR impression in a specific condition.
The characteristic buzziness also has been a source of severe degradations in analysis-and-synthesis type VOCODERs.
This degradation is made worse in statistical text-to-speech systems\cite{zen2009statistical}.
Although WaveNet\cite{oord2016wavenet} effectively made this problem disappear, a flexible and general purpose excitation signal will be beneficial for interactive and compact applications.

There have been many studies for solving these quality related problems.
Multi-band excitation is useful for improving VOCODED sound quality\cite{griffin1988multiband}.
However, direct mixing of pulse and colored noise still cannot solve the ``buzzy'' impression problem.
Multi-pulse excitation and CELP\cite{atal1982new,schroeder1985code} are also effective for reducing the ``buzzy'' impression and voice quality enhancement.
However, it is not easy to design appropriate multi-pulse for parameter manipulation, which is necessary for VOCODER-based speech conversion.

One efficient method for reducing the ``buzzy'' impression is to randomize the phase.
Group delay manipulation used in legacy-STRAIGHT was successful for reducing this impression~\cite{kawahara1999spcom}.
The log domain pulse model (LDPM) also uses phase manipulation\cite{degottex2017ieeetr}.
However, such manipulation results smearing of the signal in the time domain.
Although time windowing solves this smearing problem, it introduces other problem, power spectral modification
due to the statistical fluctuation of the truncated signal.
An element signal of the proposed FVN (Frequency domain Velvet Noise) has an excellent time-frequency localization made possible by
a six-term cosine series introduced for antialiasing glottal excitation models~\cite{kawahara2017interspeechGS}.
Also, because it is an all-pass filter's impulse response, it is
free from statistical fluctuation in power spectrum of the processed signal.

The primary focus of this article is to propose FVN and to introduce its prospective applications.
We are planning objective and subjective evaluation of FVN in various applications for the next step.
Organization of this article is as follows:
The next section briefly introduces the original velvet noise.
The following section discusses phase modification using shaping functions
which are localized both in the time and the frequency domain.
Then, applying similar procedure used in the velvet noise to
the phase modification introduced in the preceding section to define FVN.
The following section discusses three aspects of FVN useful for applications
and introduces several representative examples.
Finally, we discuss on prospective applications in speech processing
as well as application to fundamental research on human auditory processing.

\section{Velvet noise}
The velvet noise was designed for artificial reverberation algorithms.
It is a randomly allocated unit impulse sequence with minimal impulse density vs.
maximal smoothness of the noise-like characteristics.
Because such sequence can sound smoother than the Gaussian noise,
it is named ``velvet noise.''\cite{jarvelainen2007reverberation}

The velvet noise allocates a randomly selected positive or negative unit pulse at
a random location in each temporal segment\cite{jarvelainen2007reverberation,valimaki2013ieetr}.
Let $T_d$ represent the average pulse interval in samples.
The following equation determines the location of the $m$-th pulse $k_\mathrm{ovn}(m)$.
The subscript ``ovn'' stands for ``Original Velvet Noise.''
It uses two sequences of random numbers $r_1(m),$  and $r_2(m)$ generated from a uniform distribution in $(0, 1)$.
\begin{align}
k_\mathrm{ovn}(m) & = ||m Td + r_1(m) (T_d - 1) || ,
\end{align}
where the rounding function $|| \bullet ||$ returns the nearest integer.
The  following equation determines the value of the signal $s_\mathrm{ovn}(n)$ at discrete time $n$.
\begin{align}\label{eq:sigsin}
s_\mathrm{ovn}(n) & = \left\{
\begin{array}{ll}
2 || r_2(m)|| - 1  & n=  k_\mathrm{ovn}(m)  \\
0  &   \mbox{otherwise}   
\end{array}
\right. .
\end{align}

With the pulse density higher than 3,000 pulses per second, OVN sounds like a white Gaussian noise
and provides a smoother impression.
Supplemental media consists of OVN examples.

\section{Frequency domain variant of velvet noise}
The discrete Fourier transform of a velvet noise sequence closely approximates a complex Gaussian random sequence.
The discrete Fourier transform of the filtered velvet noise provides a complex Gaussian noise on the frequency axis with the filter shape weighting.
Using the duality of the frequency and the time of Fourier transform, we apply filtered velvet noise to design phase of the all-pass filter.
The impulse response of this all-pass filter is the element of the proposed FVN.
The element has the temporally localized envelope and random waveform.
The key design issue is the shape of the function to manipulate the phase.


\subsection{Unit of phase manipulation}
We use a set of cosine series functions for manipulating the phase because it is easy to implement well-behaving localization\cite{nattall1981ieee,kawahara2017interspeechGS}.
This section investigates relations between phase manipulation and the impulse response of the corresponding all-pass filter.
Let $w_p(k, B)$ represent a phase modification function on the discrete frequency domain.
The following equation provides the complex-valued impulse response $h(n; k_c, B)$ of the all-pass filter. 
\begin{align}\label{eq:unitphase}
\!\! h(n; k_c, B) & = \! \frac{1}{K}\! \sum_{k = 0}^{K - 1} \!  \exp\!\left(\frac{2 k n \pi j}{KN} + j w_p(k- k_c, B) \right) ,
\end{align}
where $k_c$ represents the discrete center frequency, and
$B$ defines the support of $w_p(k, B)$ in the frequency domain
(i.e. $w_p(k, B) = 0$ for $|k| > B$).
The symbol of the imaginary unit is $j = \sqrt{-1}$ and
$N$ represents the number of DFT bins.

We tested four types of cosine series.
They are Hann, Blackman, Nuttall, and the six-term cosine series used in\cite{kawahara2017interspeechGS}.
The Nuttall's reference\cite{nattall1981ieee} provides a list of coefficients of the first three functions and the design procedure.
The following cosine series defines these functions.
Let define $B_w = B /M$ as nominal bandwidth.
\begin{align}
w_p(k, B) & = \sum_{m = 0}^{M} a(m) \cos\left(\frac{\pi k m}{B} \right) ,
\end{align}
where $M$ represents the highest order of the cosine series.

\begin{figure}[tbp]
\begin{center}
\includegraphics[width=0.48\hsize]{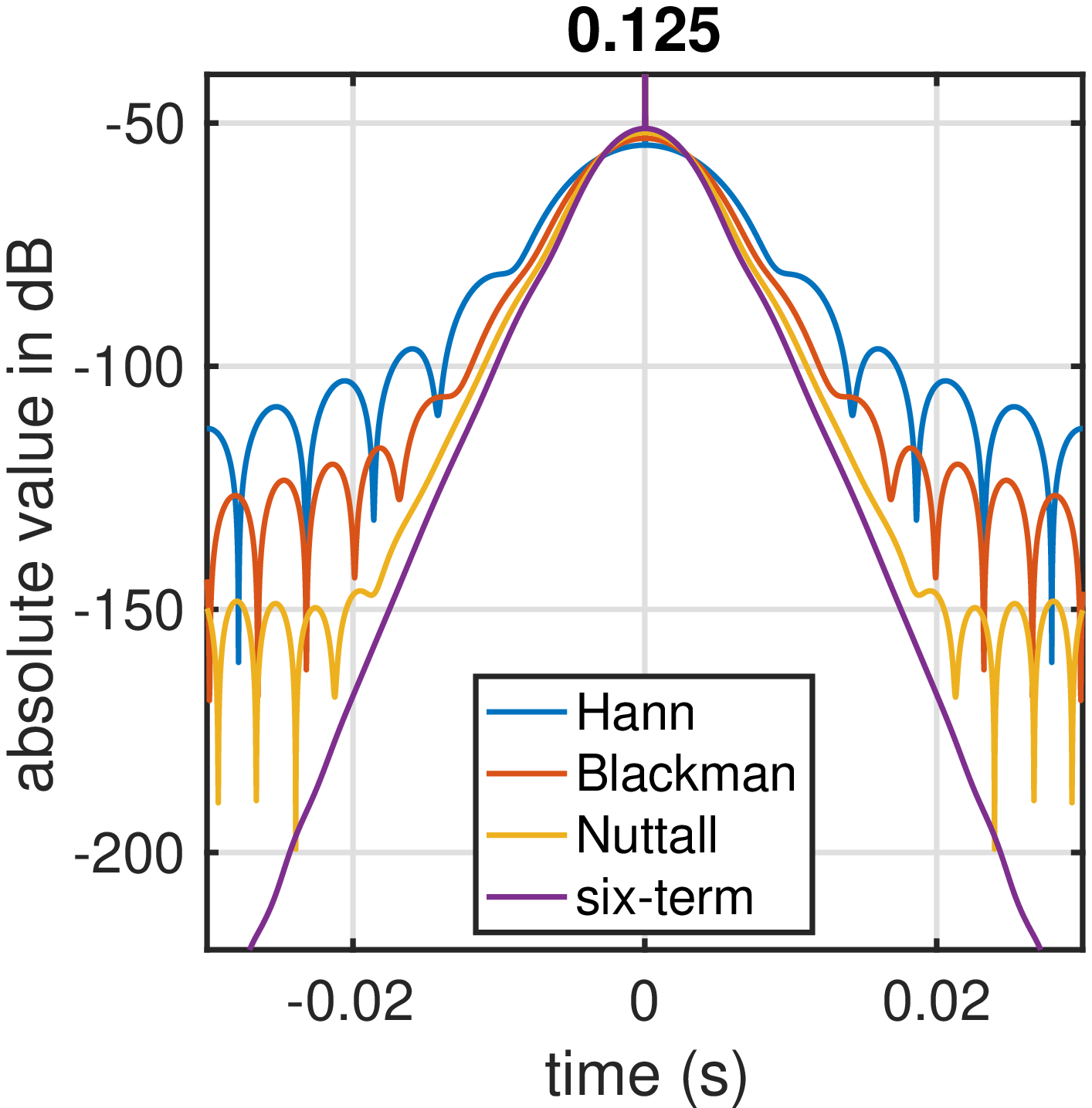}
\hfill
\includegraphics[width=0.48\hsize]{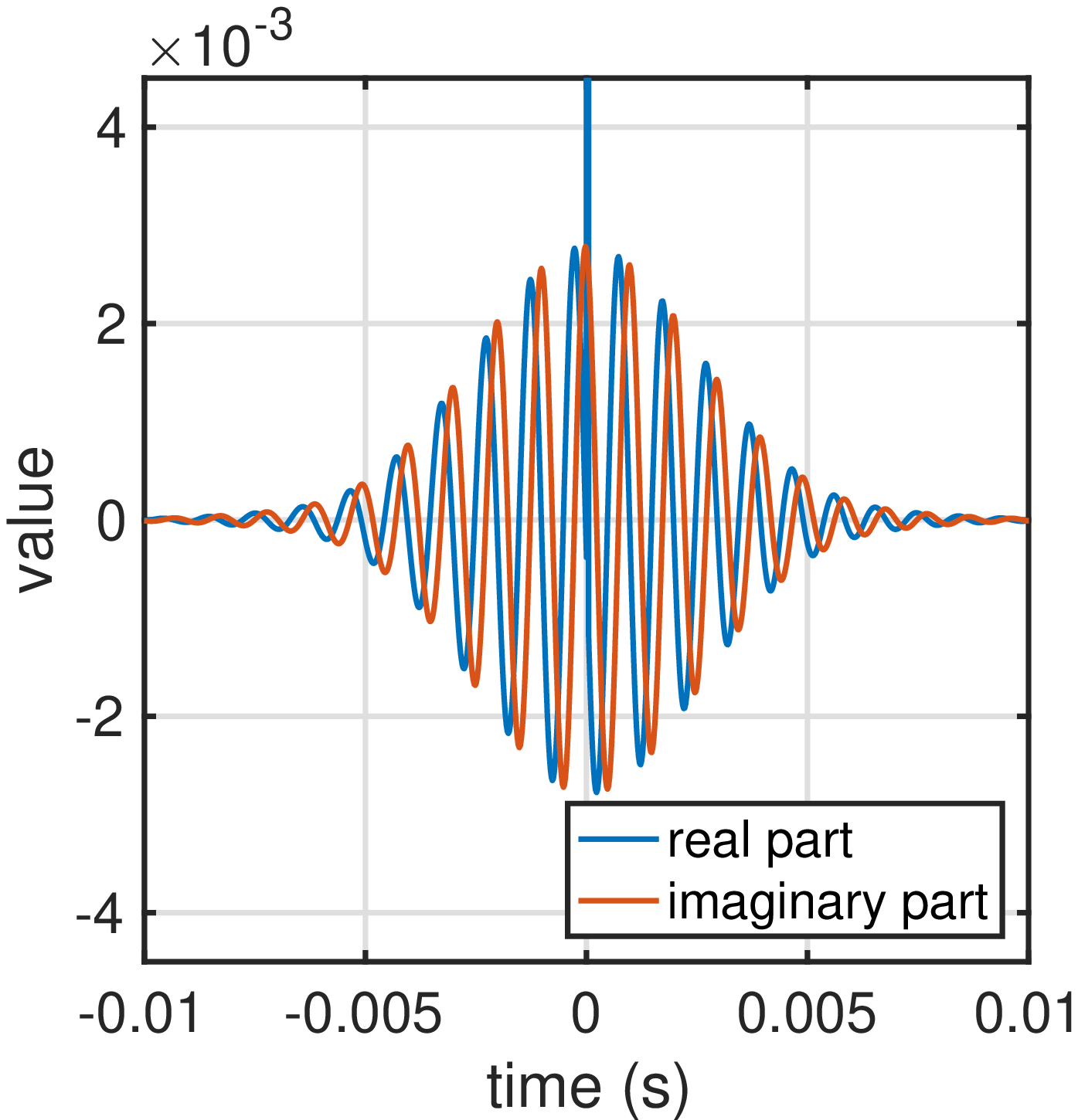}
\caption{The absolute value of the impulse response of all-pass filters made from unit phase manipulation using cosine series shapes (left plot).
An example impulse response of using the six-term series (right plot).}
\vspace{-4mm}
\label{all-passOVNk1}
\end{center}
\end{figure}
Figure~\ref{all-passOVNk1} shows examples of this phase manipulation effects.
We found that the six-term cosine series provides the best localization behavior.
The six-term series has practically no interference due to sidelobes.
We decided to use this six-term series afterward.
The coefficients of the six-term series are
0.2624710164, 0.4265335164, 0.2250165621, 0.0726831633, 0.0125124215, and 0.0007833203 from $a_0$ to $a_5$.
The sidelobes have the highest level of -114~dB and  the decay rate of -54~dB/oct (Appendix~\ref{ss:cosopt}).



\subsection{Phase manipulation unit allocation using velvet noise}
By adding unit phase manipulation $w_p(k- k_c, B)$ on
a set of center frequencies $k_c$ obeying the design rule of velvet noise yields the
filtered velvet noise in the frequency domain.
The following equation defines the allocation index (discrete frequency) $k_c = k_\mathrm{fvn}(m)$
where subscript ``fvn'' stands for Frequency domain Velvet Noise.
\begin{align}
k_\mathrm{fvn}(m) & = ||m F_d + r_1(m) (F_d - 1) || ,
\end{align}
where $F_d$ represents the average frequency segment length.
Each location spans from 0~Hz to $f_s/2$.
Let $\mathbb{K}$ represent a set of allocation indices $k_\mathrm{fvn}(m)$.
The following equation provides the phase $\varphi_\mathrm{fvn}(k)$ of this frequency variant of velvet noise.
\begin{align}
\!\! \varphi_\mathrm{fvn}(k) & = \!\!\sum_{k_c \in \mathbb{K}} \!\! s_\mathrm{fvn}(k_c) \left(w_p(k\!-\! k_c, B) \!- \!w_p(k\!+\! k_c, B)\right) , \label{eq:vpAlloc} \\
\!\! s_\mathrm{fvn}(m) & =  \left(2 || r_2(m)|| - 1 \right) \varphi_\mathrm{max} 
\end{align}
where $k$ spans discrete frequency of a DFT buffer, which has a circular discrete frequency axis and
the parameter $\varphi_\mathrm{max}$ defines the magnitude of phase manipulation.
The second term inside of parentheses of Eq.~\ref{eq:vpAlloc} is to make the phase function have the odd symmetry concerning 0~Hz and $f_s/2$.

The inverse discrete Fourier transform of this all-pass filter provides an impulse response.
It is the unit signal $h_\mathrm{fvn}(n)$ of the proposed FVN.
\begin{align}
h_\mathrm{fvn}(n) & = \frac{1}{K}\sum_{k = 0}^{K - 1}  \exp\left(\frac{2 k n \pi j }{KN} + j \varphi_\mathrm{fvn}(k) \right) .
\end{align}

\begin{figure}[tbp]
\begin{center}
\includegraphics[width=0.48\hsize]{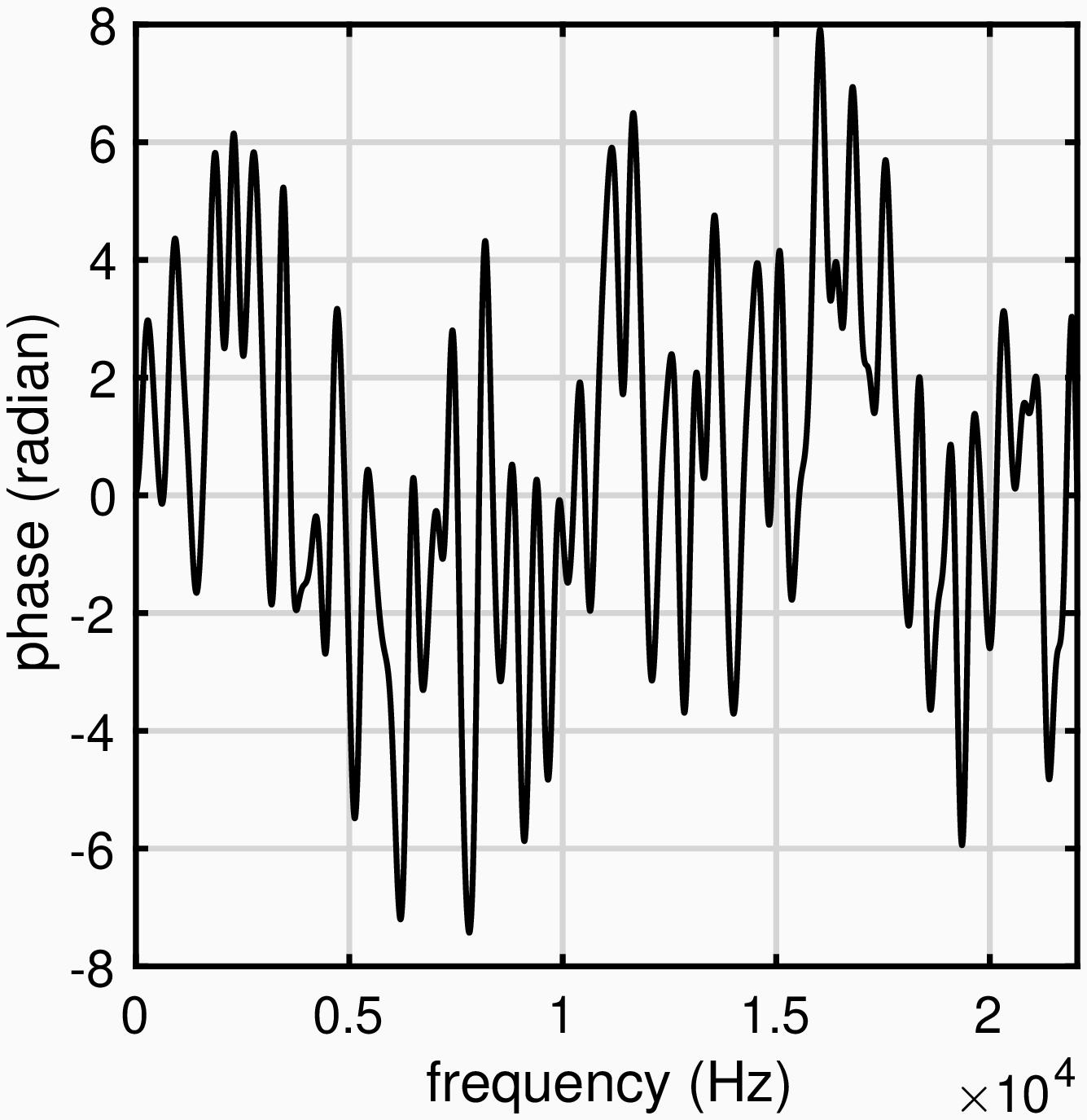}
\hfill
\includegraphics[width=0.48\hsize]{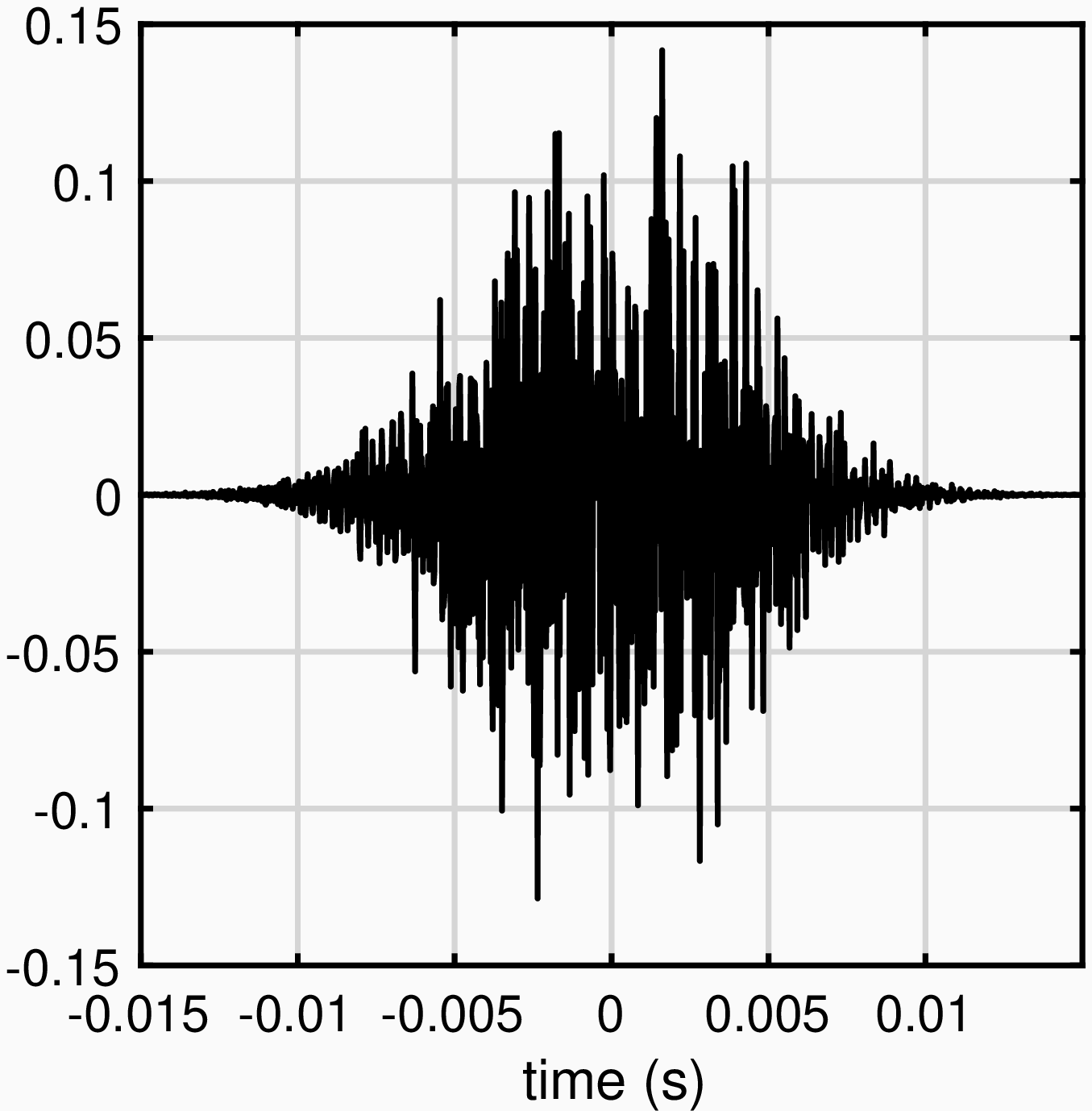}
\caption{An example of the phase (left plot) and the corresponding impulse response (right plot) of the designed all-pass filter using the six-term cosine series.}
\label{sampleFVN200wave}
\end{center}
\end{figure}
Figure~\ref{sampleFVN200wave} shows an example of the designed phase of the all-pass filter and
the corresponding impulse response.
The impulse response is temporally localized, and the phase behaves like a smoothed random sequence.
(This example uses 44,100~Hz sampling frequency, $F_d=40$~Hz, $B=200$~Hz, and $\varphi_\mathrm{max}=\pi/2$ radian.)

\begin{figure}[tbp]
\begin{center}
\includegraphics[width=0.49\hsize]{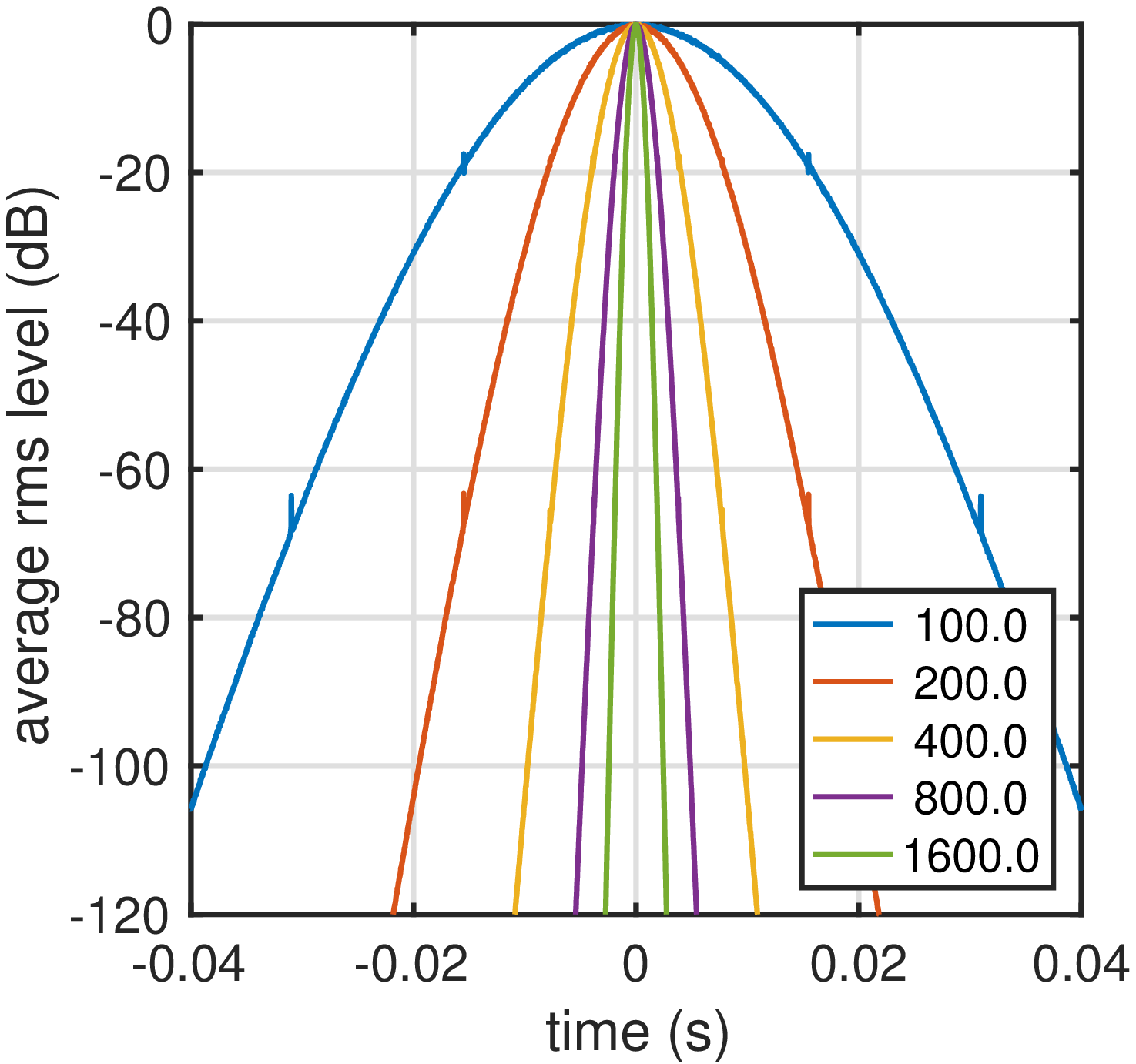}
\hfill
\includegraphics[width=0.47\hsize]{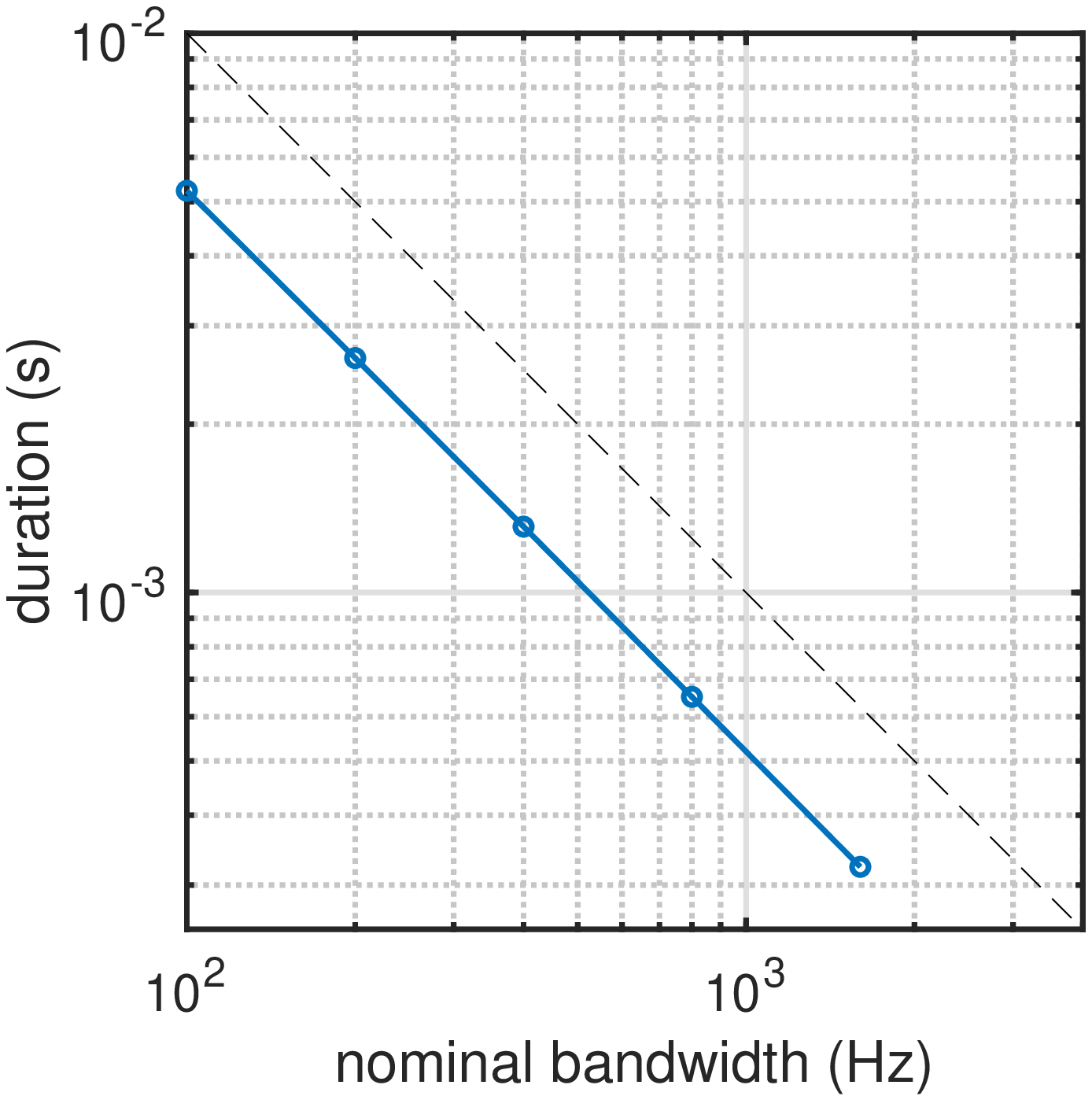}
\caption{RMS (Root Mean Squared) value of the impulse response (left plot) and
the design parameter $B$ and the duration (right plot).}
\vspace{-6mm}
\label{rmsShapeOnBw6}
\end{center}
\end{figure}
Figure~\ref{rmsShapeOnBw6} shows the simulation results of generated 5,000 FVN units.
The left plot shows the RMS (Root Mean Squared) value of the impulse response.
The legend shows the support length $B$ of the smoother $w_p(k, B)$.
The right plot shows the relation between the support length $B$ and the
duration of the impulse response.
These indicate that we can set the desired duration $\sigma_t$ of the FVN by assigning the $B$ using these simulation results.
(See Appendix~\ref{ss:durationGD} for details.)
\begin{align}
B & = \frac{0.522}{\sigma_t} \ \  \ , \ \mbox{where} \ \ \ \sigma_t^2 = \left\langle t^2 h^2_\mathrm{fvn}(t) \right\rangle
\end{align}

\subsection{Frequency dependent duration control}
FVN generated by the procedure in the previous section has a constant temporal duration in each frequency range.
It is desirable to introduce frequency-dependent temporal duration, for example, to implement voiced fricatives.
This section introduces a variant of FVN which has frequency dependent temporal duration.
We call this variant as FFVN (Frequency dependent Frequency domain Velvet Noise).

The duration of the generated FVN is proportional to the
frequency width of the smoothing function.
It suggests that by
locally warping the target frequency axis 
 implements frequency-dependent duration.
Let's define a normalized frequency weighting function $g(f) = B_\mathrm{wtgt}(f) /B_\mathrm{max}$,
where $B_\mathrm{wtgt}(f)$ represents the target duration at frequency $f$, and
$B_\mathrm{max}$ represents the maximum target duration.
The following equation defines the warped frequency axis $\nu(f)$. 
\begin{align}
\nu(f) & = \alpha \int_0^f g(u) du ,
\end{align}
where $\alpha$ is a calibration coefficient to make $\nu(f_s / 2) = f_s/2$ and
$f_s$ represents the sampling frequency.

The duration of the FVN on the morphed frequency axis $\nu$ is constant under this mapping.
The following equation provides the constant duration $B_\mathrm{worg}$ on this axis.
\begin{align}
B_\mathrm{worg} & =   B_\mathrm{wtgt}(\nu(f_\mathrm{max})) \left. \frac{d \nu(f)}{d f}\right|_{f=f_\mathrm{max}} ,
\end{align}

Mapping a constant duration FVN's phase function $\varphi_\mathrm{fvn}(\nu)$ on the new frequency axis $\nu(f)$ to the original frequency axis $f$ provides
the desired variable duration FVN's phase function $\varphi_\mathrm{mod}(f)$.
\begin{align}
\varphi_\mathrm{mod}(f) & = \varphi_\mathrm{fvn}(\nu(f)) .
\end{align}

\begin{figure}[tbp]
\begin{center}
\includegraphics[width=0.9\hsize]{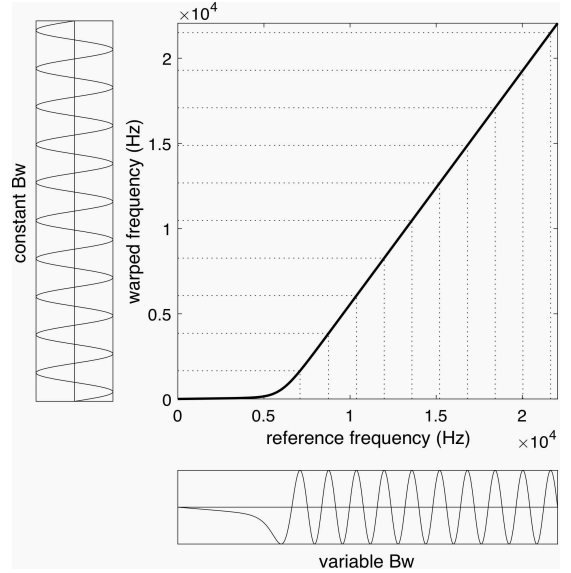}
\caption{Implementation of frequency dependent duration control using
frequency axis warping.
A generated phase function $\varphi_\mathrm{fvn}(\nu)$  shown on the left  using a constant $B$ is
converted to the modified phase function $\varphi_\mathrm{mod}(f)$ shown on the bottom, which corresponds to
the frequency dependent $B$.}
\vspace{-6mm}
\label{frequencyWarping}
\end{center}
\end{figure}
Figure~\ref{frequencyWarping} illustrates the relations between frequency axes $f$ and $\nu$,
and the phase functions $\varphi_\mathrm{mod}(f)$ and $\varphi_\mathrm{fvn}(\nu)$.
The polar form $\exp(j \varphi_\mathrm{mod}(f))$ provides the Fourier transform of the unit FVN.
This complex exponential function also is the transfer function of an all-pass filter.

The inverse discrete Fourier transform of this all-pass filter with the modified phase function $\varphi_\mathrm{mod}(f)$ provides 
the desired impulse response $h_\mathrm{mod}(n)$ of the frequency dependent duration FVN.
\begin{align}
h_\mathrm{mod}(n) & = \frac{1}{K}\sum_{k = 0}^{K - 1}  \exp\left(\frac{2 k n \pi j }{KN} + j \varphi_\mathrm{mod}(k) \right) .
\end{align}

\subsubsection{Implementation}
For defining this nonlinear mapping, we introduced a sigmoidal model and a band-wise model.
The sigmoidal model $B_\mathrm{sigm}(f)$ has two parameters, the transition frequency $f_c$ and transition width $f_{tr}$.
The band-wise model $B_\mathrm{band}(f)$ has two set of parameters, the boundary frequencies $\boldmath{f}_b(k)$ 
$(f_b(0)=0, \ldots, f_b(K)= f_s/2)$ and the
durations $\boldmath{B}_k, (k=1, \ldots, K)$ in the bands. 
It also has the additional parameter, the width $f_w$ of a raised cosine smoother $s(f; f_w) = (1 + \cos(2 \pi f/f_w))/2$ with
its support $[-f_w/2, f_w/2]$.
\begin{align}
B_\mathrm{sigm}(f) & = \frac{1}{1 + \exp\left(-\frac{f - f_c}{f_{tr}} \right)} \label{eq:sgmoidmodel} \\
B_\mathrm{band}(f) & = s(f; f_w) \ast \sum_{k=1}^{K} B_k (u_{k-1}(f) - u_k(f)) ,
\end{align}
where ``$\ast$'' represents convolution and $u_k(f)$ represents the unit step function
starting at $f_b(k)$. (See Appendix~\ref{ss:ffvnExample} for examples)

\section{Application}
This section introduces two applications of an element of FVN and FFVN.
The first one is for post-processing of VOCODER output, and
the other is data augmentation for training, for example, WaveNet.
We also introduce
the application of FVN and FFVN to the excitation source of synthetic speech.

\subsection{Time invariant all-pass filter}
Each element waveform of FVN and FFVN is an impulse response of an all-pass filter.
Applying this all-pass filter to the VOCODER outputs reduces their ``buzzy'' impression significantly.
This filtering is simple and effective post-processing for improving the quality of VOCODER outputs.
Supplemental media files provide demonstrations of  this ``buzzy'' impression reduction.

Applying this all-pass filter to an original speech sample alters the waveform significantly.
However the original and the filtered speech sound similar when the duration of the
filter is small (for example about 1~ms: consistent with\cite{Blauert1978}).
This insensitivity suggests that FVN and FFVN can be useful for data-augmentation for training, for example, WaveNet to
embed constraints due to human auditory perception.

\subsection{Excitation of synthetic speech}
Linear interpolation of the phase of two FVN or FFVN elements morphs the generated signal seamlessly.
A regularly repeated sequence of an identical FVN or FFVN element, in other words using frozen element, provides clear pitch perception.
When the elements are updated using different random number always, and the duration of each element is
longer than the repetition period, the sequence sounds like white noise.
Linear interpolation of the phase of the elements of these sequences provides an excitation signal which spans from noise to
periodic sounds seamlessly.
Supplemental media files provide demonstrations of this morphed excitation signal.

The other application of FVN and FFVN is for additive noise component of the excitation source.
Allocating an element with relatively short duration (for example, shorter than 5~ms) in each
excitation pitch period implements the temporal variation of the random component.
This implementation is effective for synthesizing low pitched voices, such as males'. 
Supplemental media files also provide demonstrations of the temporal variation of the random component.
Instead of using a Gaussian noise, substituting it with FVN and FFVN in statistical signal processing\cite{zen2009statistical} is an exciting possibility.
Using them in a complex cepstrum-based excitation model\cite{maia2012icassp} is such a prospective example.

\subsection{Supplemental media files}
Supplemental media files of this article consists of the OVN samples, and FVN and FFVN application examples.
The VOCODED speech example uses the file ``\texttt{slt/arctic\_b0436.wav}'' of CMU Arctic database\cite{kominek2004cmu}
and synthesized using Mel-Cepstrum processed envelope.
It also consists of the link to MATLAB script and resources\cite{IS2018resourcePage}.

\section{Discussion and related future work}

The proposed variants have wide potentials in applications.
High-quality and wide frequency-range recorded speech sounds have very high kurtosis in amplitude distribution\cite{kawahara2010simplification}.
The proposed all-pass filter reduces the kurtosis of the filtered signal significantly.
Since the FVN unit response is one specific type of TSP (Time Stretched Pulse),
the convolution of the processed signal with the time-reversed version recovers the original version and consequently the high kurtosis level.
This recovery is a useful feature for information hiding and tampering detection\cite{jayaram2011information}.
(See Appendix~\ref{ss:hiding}.)

OVN and FVN also provide a set of tools for investigating perceptually equivalent timbre class and fundamental properties of the human auditory system.
Effects of phase on timbre were well known\cite{plomp1969effect}, but the structure of phase-related timber was only partially investigated\cite{patterson1987pulse}.
The flexibility of FVN parameter design will open a new systematic research paradigm in auditory processing of signal phase and will provide means to revisit fundamental questions.
The apparent contradiction between auditory evoked potential and perception of the optimized chirp signal~\cite{dau2000auditory,uppenkamp2001effects}
is an example of such questions.
The answer to the question will provide the fundamental solution to the ``buzzy'' impression of VOCODED sounds.
OVN also is useful.
For example, the memory span of so-called echoic memory, around 1 to 2 seconds, was tested using random signals\cite{guttman1963lower}.
Testing periodicity perception using very sparse repeated OVN samples will shed new light on their information representation and processing.
The sparse repeated OVN may serve as a complement test signal to the IRN (Iterated Rippled Noise)\cite{yost1996jasa} used in psychoacoustic experiments.
(See Appendix~\ref{ss:ifvn}.)

It is crucially important to design systems dealing with human speech communication based on
fundamental understanding of human auditory perception mechanisms and their functions~\cite{lyon_2017}
because the end-users of such systems are humans.
Data augmentation by introducing physically different while perceptually equivalent preprocessed speech
will be one feasible strategy to introduce such understanding built into end-to-end speech applications~\cite{oord2016wavenet,2017arXiv171110433V,2017arXiv171205884S}.
All-pass filtering based on FVN provides a prospective tool for required data augmentation.

\vspace{-2mm}
\section{Conclusions}

We introduced a flexible excitation source signal which spans from a periodic signal to random signal seamlessly 
and an all-pass filter which substantially reduces ``buzzy'' impression of VOCODER outputs.
Combination of the well-behaving phase manipulation function and
the velvet noise generation procedure in the frequency domain made these important contributions possible.
We are planning to introduce this excitation source signal to reformulate perceptually isomorphic VOCODER framework.
We also make software of FVN variants and reference applications available as an open-source package on GitHub.

\vspace{-2mm}
\section{Acknowledgements}
This work was supported by JSPS KAKENHI (grants in aids for scientific research) Grant Numbers~JP15H03207, JP15H02726 and JP16K12464.

\bibliographystyle{IEEEtran}

\bibliography{IS2018Kawahara}

\begin{thebibliography}{10}
\providecommand{\url}[1]{#1}
\csname url@samestyle\endcsname
\providecommand{\newblock}{\relax}
\providecommand{\bibinfo}[2]{#2}
\providecommand{\BIBentrySTDinterwordspacing}{\spaceskip=0pt\relax}
\providecommand{\BIBentryALTinterwordstretchfactor}{4}
\providecommand{\BIBentryALTinterwordspacing}{\spaceskip=\fontdimen2\font plus
\BIBentryALTinterwordstretchfactor\fontdimen3\font minus
  \fontdimen4\font\relax}
\providecommand{\BIBforeignlanguage}[2]{{%
\expandafter\ifx\csname l@#1\endcsname\relax
\typeout{** WARNING: IEEEtran.bst: No hyphenation pattern has been}%
\typeout{** loaded for the language `#1'. Using the pattern for}%
\typeout{** the default language instead.}%
\else
\language=\csname l@#1\endcsname
\fi
#2}}
\providecommand{\BIBdecl}{\relax}
\BIBdecl

\bibitem{jarvelainen2007reverberation}
H.~J{\"a}rvel{\"a}inen and M.~Karjalainen, ``Reverberation modeling using
  velvet noise,'' in \emph{AES 30th International Conference, Saariselk{\"a},
  Finland}.\hskip 1em plus 0.5em minus 0.4em\relax Audio Engineering Society,,
  2007, pp. 15--17.

\bibitem{valimaki2013ieetr}
V.~V\"alim\"aki, H.~M. Lehtonen, and M.~Takanen, ``A perceptual study on velvet
  noise and its variants at different pulse densities,'' \emph{IEEE
  Transactions on Audio, Speech, and Language Processing}, vol.~21, no.~7, pp.
  1481--1488, July 2013.

\bibitem{kawahara2018sigmus118}
H.~Kawahara, ``Application of the velvet noise and its variant for synthetic
  speech and singing,'' \emph{IPSJ SIG Technical Report}, vol. 2018-MUS-118,
  no.~3, 2018.

\bibitem{kawahara1999spcom}
H.~Kawahara, I.~Masuda-Katsuse, and A.~de~Cheveigne, ``{Restructuring speech
  representations using a pitch-adaptive time-frequency smoothing and an
  instantaneous-frequency-based F0 extraction},'' \emph{Speech Communication},
  vol.~27, no. 3-4, pp. 187--207, 1999.

\bibitem{Kawahara2001maveba}
H.~Kawahara, J.~Estill, and O.~Fujimura, ``{Aperiodicity extraction and control
  using mixed mode excitation and group delay manipulation for a high quality
  speech analysis, modification and synthesis system STRAIGHT},'' in
  \emph{Proceedings of MAVEBA}, Firentze Italy, 2001, pp. 59--64.

\bibitem{kawahara2010simplification}
H.~Kawahara, M.~Morise, T.~Takahashi, H.~Banno, R.~Nisimura, and T.~Irino,
  ``Simplification and extension of non-periodic excitation source
  representations for high-quality speech manipulation systems,'' in
  \emph{Interspeech 2010}, Makuhari Japan, 2010, pp. 38--41.

\bibitem{degottex2017ieeetr}
G.~Degottex, P.~Lanchantin, and M.~Gales, ``A log domain pulse model for
  parametric speech synthesis,'' \emph{IEEE/ACM Transactions on Audio, Speech,
  and Language Processing}, vol.~26, no.~1, pp. 57--70, Jan 2018.

\bibitem{kawahara2008icassp}
H.~Kawahara, M.~Morise, T.~Takahashi, R.~Nisimura, T.~Irino, and H.~Banno,
  ``{TANDEM-STRAIGHT: A temporally stable power spectral representation for
  periodic signals and applications to interference-free spectrum, F0 and
  aperiodicity estimation},'' in \emph{ICASSP 2008}, Las Vegas, 2008, pp.
  3933--3936.

\bibitem{morise2016world}
M.~Morise, F.~Yokomori, and K.~Ozawa, ``World: A vocoder-based high-quality
  speech synthesis system for real-time applications,'' \emph{IEICE
  TRANSACTIONS on Information and Systems}, vol.~99, no.~7, pp. 1877--1884,
  2016.

\bibitem{oppenheimerBook}
A.~V. Oppenheim and R.~W. Schafer, \emph{Discrete-time signal processing:
  Pearson new International Edition}.\hskip 1em plus 0.5em minus 0.4em\relax
  Pearson Higher Ed., 2013.

\bibitem{yegnanarayana1998ieeetr}
B.~Yegnanarayana, C.~{d}'Alessandro, and V.~Darsinos, ``{An iterative algorithm
  for decomposition of speech signals into periodic and aperiodic
  components},'' \emph{IEEE Transactions on Speech and Audio Processing},
  vol.~6, no.~1, pp. 1--11, jan 1998.

\bibitem{Malyska2008}
N.~Malyska and T.~F. Quatieri, ``{Spectral representations of nonmodal
  phonation},'' \emph{IEEE Transactions on Audio, Speech and Language
  Processing}, vol.~16, no.~1, pp. 34--46, 2008.

\bibitem{Skoglund2000ieeetrans}
J.~Skoglund and W.~B. Kleijn, ``{On time-frequency masking in voiced speech},''
  \emph{Speech and Audio Processing, IEEE Transactions on}, vol.~8, no.~4, pp.
  361--369, 2000.

\bibitem{zen2009statistical}
H.~Zen, K.~Tokuda, and A.~W. Black, ``Statistical parametric speech
  synthesis,'' \emph{Speech Communication}, vol.~51, no.~11, pp. 1039--1064,
  2009.

\bibitem{oord2016wavenet}
A.~{van~den~Oord}, S.~Dieleman, H.~Zen, K.~Simonyan, O.~Vinyals, A.~Graves,
  N.~Kalchbrenner, A.~Senior, and K.~Kavukcuoglu, ``{WaveNet: A generative
  model for raw audio},'' \emph{arXiv preprint arXiv:1609.03499}, pp. 1--15,
  2016.

\bibitem{griffin1988multiband}
D.~W. Griffin and J.~S. Lim, ``Multiband excitation vocoder,'' \emph{IEEE
  Transactions on acoustics, speech, and signal processing}, vol.~36, no.~8,
  pp. 1223--1235, 1988.

\bibitem{atal1982new}
B.~Atal and J.~Remde, ``A new model of lpc excitation for producing
  natural-sounding speech at low bit rates,'' in \emph{Acoustics, Speech, and
  Signal Processing, IEEE International Conference on ICASSP'82.},
  vol.~7.\hskip 1em plus 0.5em minus 0.4em\relax IEEE, 1982, pp. 614--617.

\bibitem{schroeder1985code}
M.~Schroeder and B.~Atal, ``Code-excited linear prediction (celp): High-quality
  speech at very low bit rates,'' in \emph{Acoustics, Speech, and Signal
  Processing, IEEE International Conference on ICASSP'85.}, vol.~10.\hskip 1em
  plus 0.5em minus 0.4em\relax IEEE, 1985, pp. 937--940.

\bibitem{kawahara2017interspeechGS}
H.~Kawahara, K.-I. Sakakibara, M.~Morise, H.~Banno, T.~Toda, and T.~Irino, ``{A
  new cosine series antialiasing function and its application to aliasing-free
  glottal source models for speech and singing synthesis},'' in \emph{Proc.
  Interspeech 2017}, Stocholm, August 2017, pp. 1358--1362.

\bibitem{nattall1981ieee}
A.~H. Nuttall, ``{Some windows with very good sidelobe behavior},'' \emph{IEEE
  Trans. Audio Speech and Signal Processing}, vol.~29, no.~1, pp. 84--91, 1981.

\bibitem{Blauert1978}
J.~Blauert and P.~Laws, ``{Group delay distortions in electroacoustical
  systems},'' \emph{Journal of the Acoustical Society of America}, vol.~63,
  no.~5, pp. 1478--1483, 1978.

\bibitem{maia2012icassp}
R.~Maia, M.~Akamine, and M.~J.~F. Gales, ``Complex cepstrum as phase
  information in statistical parametric speech synthesis,'' in \emph{2012 IEEE
  International Conference on Acoustics, Speech and Signal Processing
  (ICASSP)}, March 2012, pp. 4581--4584.

\bibitem{kominek2004cmu}
J.~Kominek and A.~W. Black, ``The {CMU Arctic} speech databases,'' in
  \emph{Fifth ISCA Workshop on Speech Synthesis}, 2004.

\bibitem{IS2018resourcePage}
\BIBentryALTinterwordspacing
H.~Kawahara, ``{Resource page for Interspeech 2018},'' (Last access:
  10/June/2018). [Online]. Available:
  \url{http://www.wakayama-u.ac.jp/\%7ekawahara/IS2018/}
\BIBentrySTDinterwordspacing

\bibitem{jayaram2011information}
P.~Jayaram, H.~Ranganatha, and H.~Anupama, ``Information hiding using audio
  steganography--a survey,'' \emph{The International Journal of Multimedia \&
  Its Applications (IJMA) Vol}, vol.~3, pp. 86--96, 2011.

\bibitem{plomp1969effect}
R.~Plomp and H.~J. Steeneken, ``Effect of phase on the timbre of complex
  tones,'' \emph{The Journal of the Acoustical Society of America}, vol.~46,
  no.~2B, pp. 409--421, 1969.

\bibitem{patterson1987pulse}
R.~D. Patterson, ``A pulse ribbon model of monaural phase perception,''
  \emph{The Journal of the Acoustical Society of America}, vol.~82, no.~5, pp.
  1560--1586, 1987.

\bibitem{dau2000auditory}
T.~Dau, O.~Wegner, V.~Mellert, and B.~Kollmeier, ``Auditory brainstem responses
  with optimized chirp signals compensating basilar-membrane dispersion,''
  \emph{The Journal of the Acoustical Society of America}, vol. 107, no.~3, pp.
  1530--1540, 2000.

\bibitem{uppenkamp2001effects}
S.~Uppenkamp, S.~Fobel, and R.~D. Patterson, ``The effects of temporal
  asymmetry on the detection and perception of short chirps,'' \emph{Hearing
  research}, vol. 158, no. 1-2, pp. 71--83, 2001.

\bibitem{guttman1963lower}
N.~Guttman and B.~Julesz, ``Lower limits of auditory periodicity analysis,''
  \emph{The Journal of the Acoustical Society of America}, vol.~35, no.~4, pp.
  610--610, 1963.

\bibitem{yost1996jasa}
W.~A. Yost, R.~Patterson, and S.~Sheft, ``A time domain description for the
  pitch strength of iterated rippled noise,'' \emph{The Journal of the
  Acoustical Society of America}, vol.~99, no.~2, pp. 1066--1078, 1996.

\bibitem{lyon_2017}
R.~F. Lyon, \emph{Human and Machine Hearing: Extracting Meaning from
  Sound}.\hskip 1em plus 0.5em minus 0.4em\relax Cambridge University Press,
  2017.

\bibitem{2017arXiv171110433V}
A.~{van den Oord}, Y.~{Li}, I.~{Babuschkin}, K.~{Simonyan}, O.~{Vinyals},
  K.~{Kavukcuoglu}, G.~{van den Driessche}, E.~{Lockhart}, L.~C. {Cobo},
  F.~{Stimberg}, N.~{Casagrande}, D.~{Grewe}, S.~{Noury}, S.~{Dieleman},
  E.~{Elsen}, N.~{Kalchbrenner}, H.~{Zen}, A.~{Graves}, H.~{King},
  T.~{Walters}, D.~{Belov}, and D.~{Hassabis}, ``{Parallel WaveNet: Fast
  High-Fidelity Speech Synthesis},'' \emph{ArXiv e-prints}, Nov. 2017.

\bibitem{2017arXiv171205884S}
J.~{Shen}, R.~{Pang}, R.~J. {Weiss}, M.~{Schuster}, N.~{Jaitly}, Z.~{Yang},
  Z.~{Chen}, Y.~{Zhang}, Y.~{Wang}, R.~{Skerry-Ryan}, R.~A. {Saurous},
  Y.~{Agiomyrgiannakis}, and Y.~{Wu}, ``{Natural TTS Synthesis by Conditioning
  WaveNet on Mel Spectrogram Predictions},'' \emph{ArXiv e-prints}, Dec. 2017.

\bibitem{fujisaki1986icassp}
H.~Fujisaki and M.~Ljungqvist, ``{Proposal and evaluation of models for the
  glottal source waveform},'' in \emph{ICASSP 1986}, Tokyo, 1986, pp.
  1605--1608.

\bibitem{fujisaki1987icassp}
------, ``{Estimation of voice source and vocal tract parameters based on ARMA
  analysis and a model for the glottal source waveform},'' in \emph{ICASSP
  1987}, 1987, pp. 637--640.

\bibitem{fant1985four}
G.~Fant, J.~Liljencrants, and Q.-g. Lin, ``{A four-parameter model of glottal
  flow},'' \emph{STL-QPSR}, vol.~4, no. 1985, pp. 1--13, 1985.

\bibitem{Kawahara2017interspeechMod}
H.~Kawahara, K.-I. Sakakibara, M.~Morise, H.~Banno, and T.~Toda, ``{A
  modulation property of time-frequency derivatives of filtered phase and its
  application to aperiodicity and fo estimation},'' in \emph{Proc. Interspeech
  2017}, 2017, pp. 424--428.

\bibitem{kawahara2017pitfall}
H.~Kawahara, ``Pitfalls in digital signal processing,'' \emph{The Journal of
  the Acoustical Society of Japan}, vol.~73, no.~9, pp. 592--599, 2017, [in
  Japanese].

\bibitem{kaiser1980use}
J.~Kaiser and R.~W. Schafer, ``{On the use of the $I_0$-sinh window for
  spectrum analysis},'' \emph{Acoustics, Speech and Signal Processing, IEEE
  Transactions on}, vol.~28, no.~1, pp. 105--107, 1980.

\bibitem{slepian1961prolate}
D.~Slepian and H.~O. Pollak, ``{Prolate spheroidal wave functions, Fourier
  analysis and uncertainty-I},'' \emph{Bell System Technical Journal}, vol.~40,
  no.~1, pp. 43--63, 1961.

\bibitem{cohen95}
L.~Cohen, \emph{{Time-frequency analysis}}.\hskip 1em plus 0.5em minus
  0.4em\relax Englewood Cliffs, NJ: Prentice Hall, 1995.

\bibitem{zen2007details}
H.~Zen, T.~Toda, M.~Nakamura, and K.~Tokuda, ``Details of the {Nitech
  HMM-based} speech synthesis system for the {Blizzard Challenge} 2005,''
  \emph{IEICE transactions on information and systems}, vol.~90, no.~1, pp.
  325--333, 2007.

\bibitem{Kawahara2010intrsp}
H.~Kawahara, M.~Morise, T.~Takahashi, H.~Banno, R.~Nisimura, and T.~Irino,
  ``{Simplification and extension of non-periodic excitation source
  representations for high-quality speech manipulation systems},'' in
  \emph{Proc. Interspeech 2010}, no. September, Tokyo, 2010, pp. 38--41.

\bibitem{Kominek2004}
J.~Kominek and A.~Black, ``The {CMU} {A}rctic databases for speech synthesis,''
  \emph{Proc. {ISCA} Workshop on Speech Synthesis}, pp. 223--224, 2004.

\bibitem{Childers1991}
D.~G. Childers and C.~K. Lee, ``{Vocal quality factors: analysis, synthesis,
  and perception.}'' \emph{The Journal of the Acoustical Society of America},
  vol.~90, pp. 2394--2410, 1991.

\end{thebibliography}


\appendix
\section{Six-term cosine series}\label{ss:cosopt}
The phase manipulation function used in this article is a six-term cosine series optimized to yield the lowest sidelobe level with the steepest possible sidelobe decay rate using
the procedure introduced in\cite{nattall1981ieee}.
In the optimization procedure, we introduced constraints that up to the third order, the derivatives at the ends of support of the function are zero.
These constraints made the procedure a single-parameter optimization.
We conducted an exhaustive search using the truncated representation (ten-digits in the right of the decimal point)
 in the vicinity of the parameter optimized using the floating-point representation of MATLAB.
The coefficients:
0.2624710164, 0.4265335164, 0.2250165621, 0.0726831633, 0.0125124215, and 0.0007833203 from zeroth to fifth order coefficients
are the result of this exhaustive search.

\begin{figure}[tbp]
\begin{center}
\begin{center}
\includegraphics[width=75mm]{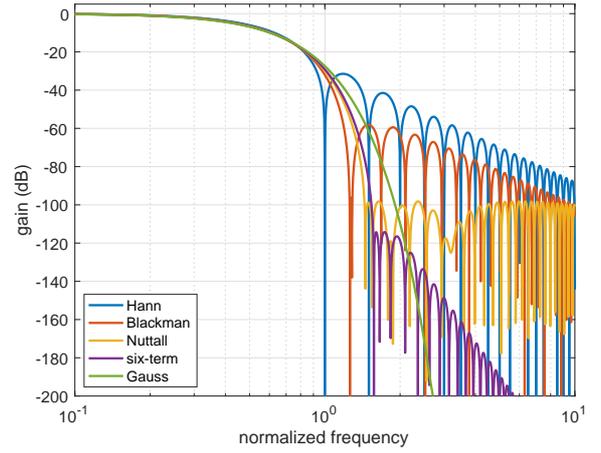}
\end{center}
\caption{Gain of cosine series. The green line shows the reference Gaussian function.
Duration of all windows are adjusted to match the duration of Hann window.}
\vspace{-4mm}
\label{cosSerWinGain}
\end{center}
\end{figure}
Figure~\ref{cosSerWinGain} shows gains of cosine series functions and the reference Gaussian function.
The duration defined using the second moment of all windows are adjusted to have the same length.

The original use of this six-term cosine series was for antialiasing Fujisaki-Ljungqvist model\cite{fujisaki1986icassp,fujisaki1987icassp} 
and L-F glottal source model\cite{fant1985four} in closed-form equations\cite{kawahara2017interspeechGS}.
The function turned out to be the best function for simultaneous SNR and $f_\mathrm{o}$ estimation\cite{Kawahara2017interspeechMod}.
It is because the optimized six-term cosine series does not introduce glitches in instantaneous frequency calculation while other commonly used functions suffer\cite{kawahara2017pitfall}.
The functions tested were
Hamming, Blackman, Nuttall (function \texttt{nuttallwin} in MATLAB), Kaiser\cite{kaiser1980use} and 
the prolate spheroidal wave function\cite{slepian1961prolate} (function \texttt{dpss} in MATLAB).
For Kaiser and the \texttt{dpss}, their parameters were adjusted to have the closest maximum sidelobe level
to that of Nuttall's.

\section{FVN: duration, ERL, and group delay}\label{ss:durationGD}
\begin{figure}[tbp]
\begin{center}
\begin{center}
\includegraphics[width=75mm]{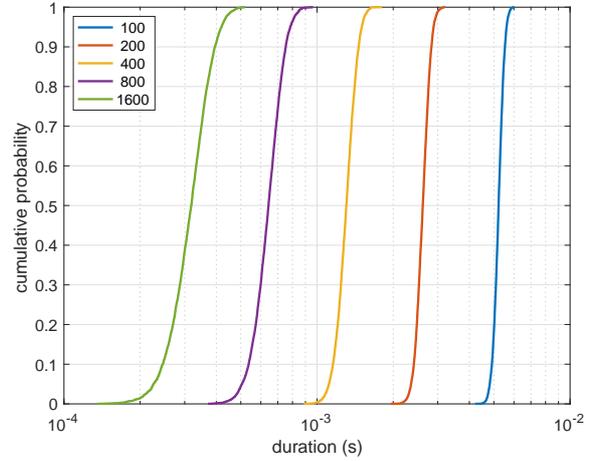}
\end{center}
\caption{Cumulative distribution of the generated FVNs.
The legend shows the support length $B$ (Hz).
The average frequency segment length uses $F_d = B / 3$ in this simulation.
The number of iteration is 5.000.}
\vspace{-4mm}
\label{cumProbDv3s}
\end{center}
\end{figure}
We use three measures to represent the length of the generated FVN.
They are duration, ERL (Effective Rectangular Length), and group delay.
This appendix describes their relations.

The duration $\sigma_t$ of a signal $x(t)$ is defined as the second order moment of the waveform in the time domain.
\begin{align}
\sigma_t & = \left( \frac{\int_D t^2 |x(t)|^2 dt}{\int_D  |x(t)|^2 dt}\right)^{\frac{1}{2}} ,
\end{align}
where $D$ represents the support of the signal $x(t)$.

The ERL $\xi$ is the normalized duration by the duration of a rectangle signal of a unit length.
\begin{align}
\xi & = \frac{\sigma_t}{\left(\int_{-1/2}^{1/2} t^2 dt\right)^{\frac{1}{2}}} \approx  3.4638 \sigma_t   .
\end{align}

The group delay $\tau_\mathrm{g}(\omega)$ of the signal $x(t)$ is defined using its Fourier transform $X(\omega) = a(\omega)\exp(j\theta(\omega))$,
where $a(\omega)$ represents the absolute value of $X(\omega)$.
\begin{align}
\tau_\mathrm{g}(\omega) & = -\frac{d \theta(\omega)}{d \omega} ,
\end{align}
where $\omega = 2 \pi f$.

Note that the power weighted average time is equal to the 
average of power spectrum weighted group delay $\sigma_\omega$.
This equation represents that the group delay at a frequency $\omega$ is the
temporal centroid of the power at the frequency.
\begin{align}
\frac{\int_D t |x(t)|^2 dt}{\int_D  |x(t)|^2 dt} & = \frac{\int \tau_\mathrm{g}(\omega) |X(\omega)|^2  d\omega}{\int |X(\omega)|^2 d\omega} . \label{eq:gdmeaning}
\end{align}

Generally the duration consists of the contribution both from amplitude and phase of its Fourier transform\cite{cohen95}.
Note that the second order moment of the group delay equals the signal duration $\sigma_t$, when $|X(\omega)|$ is constant regarding frequency.

Figure~\ref{cumProbDv3s} shows simulation results of 5,000 iterations.
The horizontal axis represents the observed duration, and the
vertical axis represents the cumulative probability.
The right plot of Fig.~\ref{rmsShapeOnBw6} in the text uses the median of each distribution.

\begin{figure}[tbp]
\begin{center}
\begin{center}
\includegraphics[width=75mm]{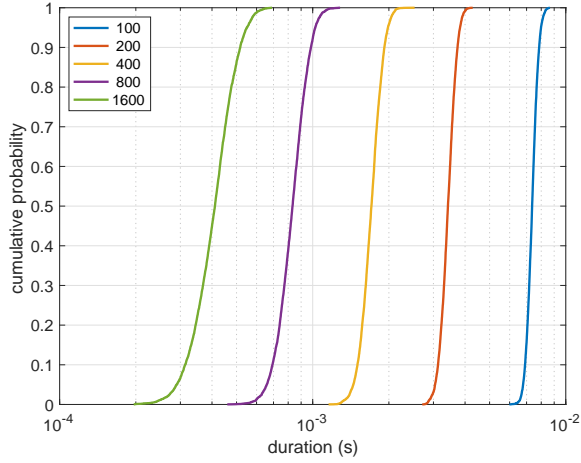}
\end{center}
\caption{Cumulative distribution of the generated FVNs.
The legend shows the support length $B$ (Hz).
The average frequency segment length uses $F_d = B / 5$ in this simulation.
The number of iteration is 5.000.}
\label{cumProbDv5s}
\end{center}
\end{figure}
Figure~\ref{cumProbDv5s} shows another simulation results.
More dense allocation $F_d = B / 5$ is used.

\section{Frequency dependent duration control: FFVN examples}\label{ss:ffvnExample}
This appendix presents two frequency shaping examples of FFVN. 
The frequency shaping functions are sigmoid and a table of frequency-duration pairs.

\subsection{Frequency shaping using sigmoid}
\begin{figure}[tbp]
\begin{center}
\begin{center}
\includegraphics[width=80mm]{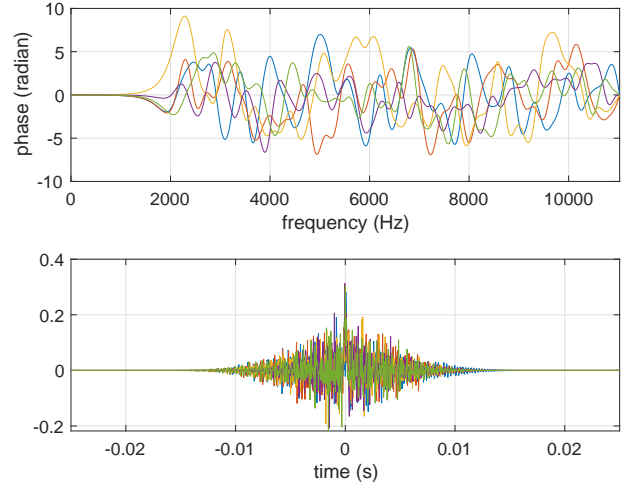}
\end{center}
\caption{Frequency weighting example using sigmoid weighting.
The upper panel shows designed phase plots of five iterations.
The lower panel shows the generated signal plots of five iterations.}
\label{ffvnSigmoidPhWv}
\end{center}
\end{figure}
Figure~\ref{ffvnSigmoidPhWv} shows examples of the FFVN with the following sigmoidal weighting in the frequency domain.
\begin{align}
B_\mathrm{wtgt}(f) & = \left(\frac{1 - c_\mathrm{floor}}{1 + \exp\left(- \frac{f - f_\mathrm{c}}{f_\mathrm{tr}}\right)} + c_\mathrm{floor}\right) B_\mathrm{max} \\
& \mbox{where} \ \ c_\mathrm{floor} = \frac{B_\mathrm{min}}{B_\mathrm{max}} , \nonumber
\end{align}
where $B_\mathrm{max}$ and $B_\mathrm{min}$ represent the maximum and minimum values of $B_\mathrm{wtgt}(f)$.
This equation corresponds to Eq.~\ref{eq:sgmoidmodel}.
The minimum value $B_\mathrm{min}$ is to prevent implementation problem.
The setting for Fig.~\ref{ffvnSigmoidPhWv} are $B_\mathrm{max} = 3$~ms and $B_\mathrm{min}=0.0037$~ms.
The corner frequency $f_c$ is 2,000~Hz, and the transition frequency parameter $f_\mathrm{tr}$ is 200~Hz.

The upper plot shows the phase of FFVN.
The plot overlays five FFVN phase samples.
The lower plot shows the waveform of FFVN.
It also overlays five FFVN waveform samples.
The sampling frequency of this simulation is $22,050$~Hz.
The length of the FFT buffer is 32768.
This unusually long FFT buffer is to reduce spurious components caused by the linear interpolation
used to implement the frequency warping.
In practical use, a buffer length longer than five to ten times of the maximum duration of FFVN is relevant.

\begin{figure}[tbp]
\begin{center}
\includegraphics[width=0.9\hsize]{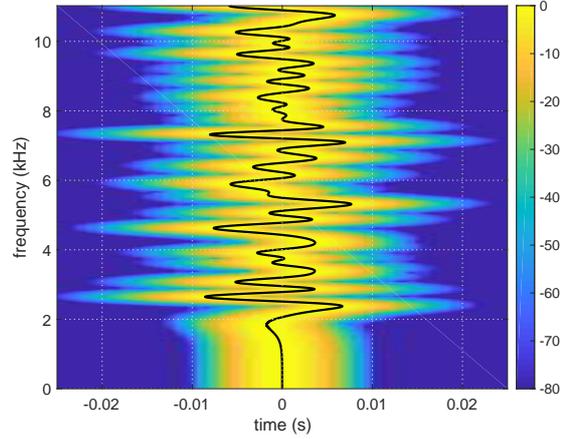}
\caption{Frequency weighting example using sigmoid weighting.
The image shows one example of the spectrogram of FFVN.
The solid black line shows the corresponding group delay.}
\vspace{-4mm}
\label{ffvnSigmoidSgramExWithGD}
\end{center}
\end{figure}
\begin{figure}[tbp]
\begin{center}
\includegraphics[width=0.9\hsize]{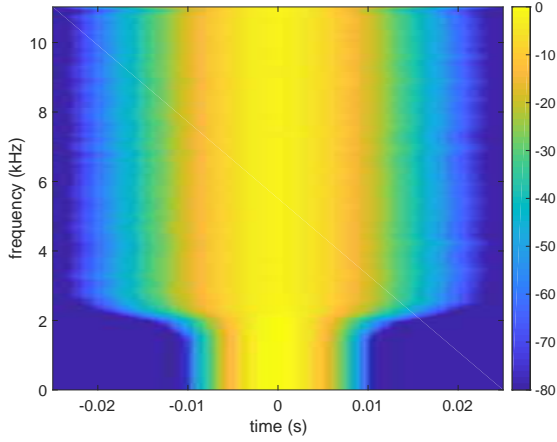}
\caption{Frequency weighting example using sigmoid weighting.
The image shows the averaged spectrogram of 5,000 FFVN samples.}
\vspace{-4mm}
\label{ffvnSigmoidSgramAve}
\end{center}
\end{figure}
Figure~\ref{ffvnSigmoidSgramExWithGD} shows an example spectrogram of FFVN.
The time windowing function used to calculate these spectrograms is \texttt{nuttallwin} of MATLAB.
It is the 15-th item in Table~II of the reference\cite{nattall1981ieee}.
The window length is 20~ms, and the frameshift is 0.5~ms.
The overlaid solid black line shows the group delay $\tau_\mathrm{gmod}(f)$ 
which is calculated from the modified phase function $\varphi_\mathrm{mod}(f)$.
Note the group delay coincides with the peak of the spectrogram at each frequency.
It is consistent with the note on Eq.~\ref{eq:gdmeaning}.

Figure~\ref{ffvnSigmoidSgramAve} shows spectrograms of FFVN examples.
The left image shows one example.
By power averaging 5,000 spectrograms, it yields the averaged spectrogram shown in the right image.
The image illustrates how the target sigmoid shapes the average spread of power.

\subsection{Frequency shaping using frequency-duration table}
For some applications a table defining the duration of each frequency band is convenient.
Table~\ref{fbtable} shows an example of such definition.
This simulation adopts the frequency division and
the sampling frequency conditions from the HTS statistical speech synthesis package\cite{zen2007details}.

\begin{table}[tbp]
\caption{Table for defining duration of each frequency band.}
\begin{center}
\begin{tabular}{c|c}
\hline 
frequency band (Hz) & duration (ms) \\ \hline
\hspace{1.6em}0 -- 1000 & 0.1 \\
1000 -- 2000 & 0.4 \\
2000 -- 4000 & 3 \\
4000 -- 6000 & 2 \\
6000 -- $f_s/2$ & 5  \\ \hline
\end{tabular}
\end{center}
\label{fbtable}
\end{table}
~
\begin{figure}[tbp]
\begin{center}
\begin{center}
\includegraphics[width=0.9\hsize]{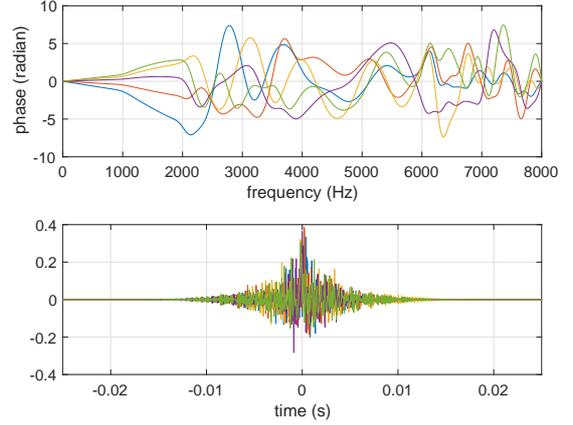}
\end{center}
\caption{Frequency weighting example of FFVN generated by using Table~\ref{fbtable}.
The upper panel shows designed phase plots of five iterations.
The lower panel shows the generated signal plots of five iterations.}
\vspace{-4mm}
\label{ffvnBandLPhWv}
\end{center}
\end{figure}
Figure~\ref{ffvnBandLPhWv} shows the phase function and the generated FFVN using Table~\ref{fbtable}.
The speed of phase change differs in each frequency band.
Shorter duration corresponds to the slower change, and the longer duration corresponds to the faster change.

\begin{figure}[tbp]
\begin{center}
\includegraphics[width=0.9\hsize]{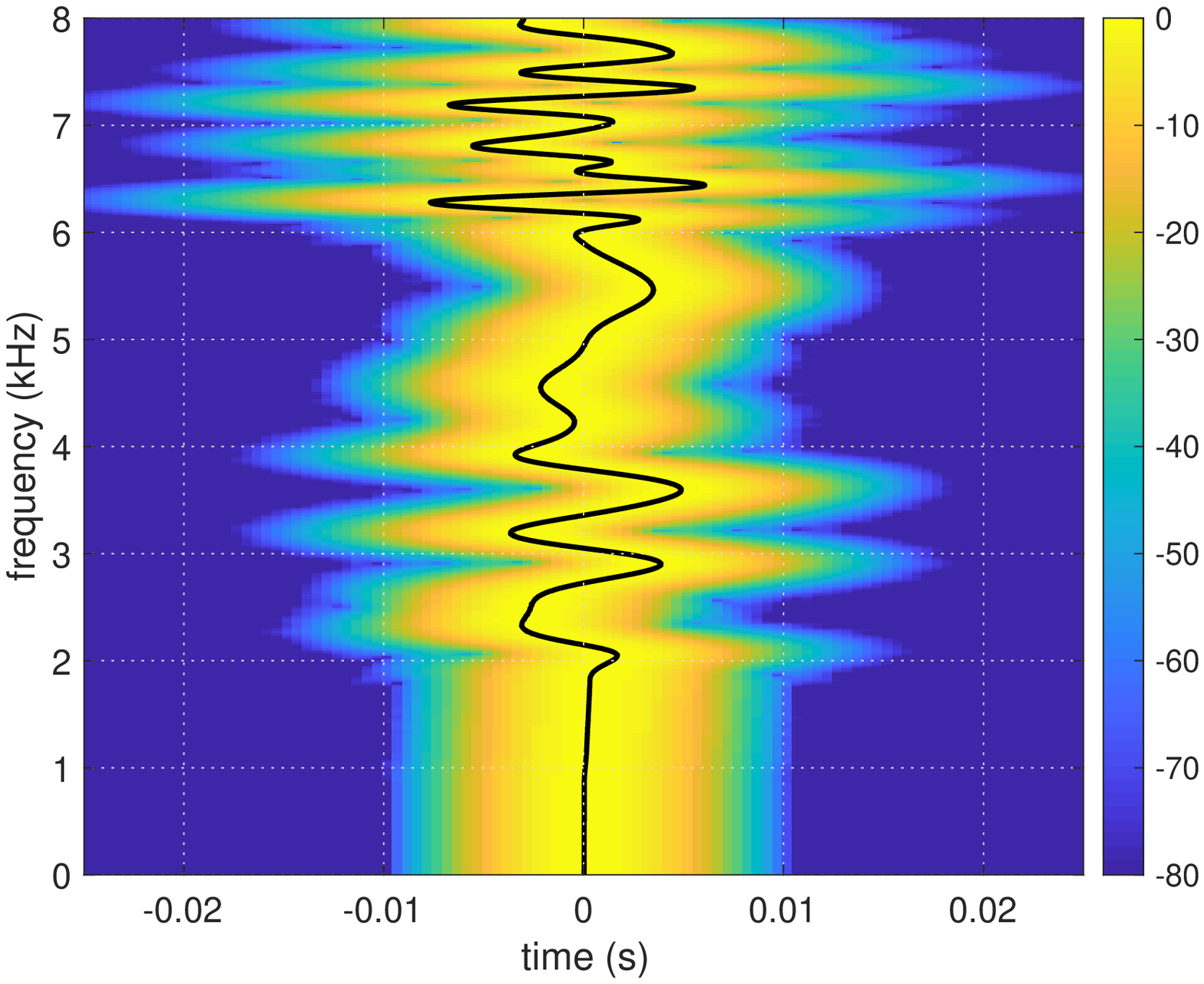}
\includegraphics[width=0.9\hsize]{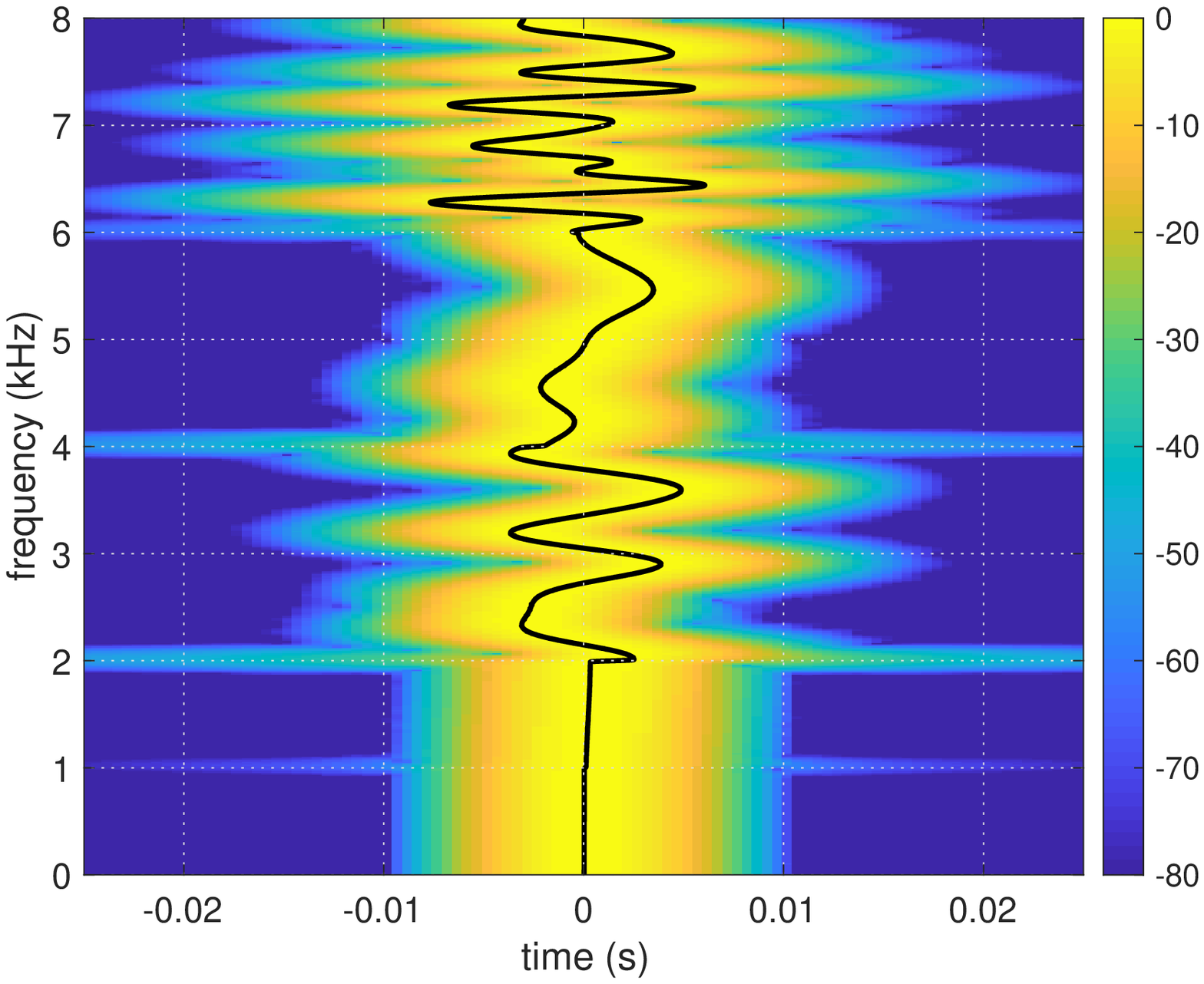}
\caption{Frequency weighting example using a frequency-band table.
The image shows two examples of the spectrogram of FFVN.
The solid black line shows the corresponding group delay.
The upper image uses $f_w=400$~Hz. The lower image uses $f_w=25$~Hz.
}
\vspace{-4mm}
\label{ffvnBandLSgramExWithGD}
\end{center}
\end{figure}
Figure~\ref{ffvnBandLSgramExWithGD} shows an example spectrogram of FFVN generated using Table~\ref{fbtable}.
The width $f_w$ of the smoother is 400~Hz for the upper image
and 25~Hz for the lower image.
Note that the lower image has power smearing at the boundary of each frequency band.
This smearing is caused by the sharp group delay slope transition at each boundary.
The overlaid solid black line shows the group delay $\tau_\mathrm{gmod}(f)$ of the modified phase function $\varphi_\mathrm{mod}(f)$.

\begin{figure}[tbp]
\begin{center}
\includegraphics[width=0.9\hsize]{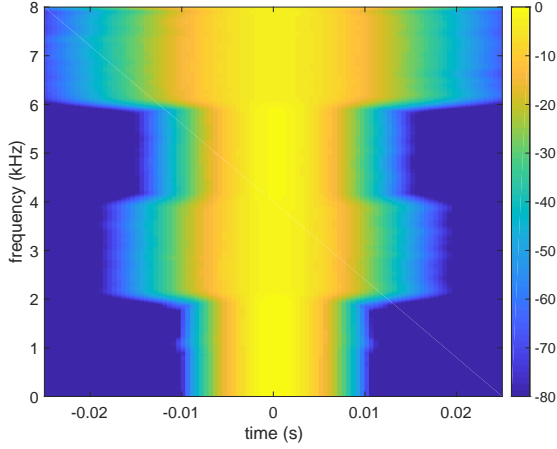}
\caption{Frequency weighting example generated by using using Table~\ref{fbtable}.
The image shows the averaged spectrogram of 5,000 FFVN samples.}
\label{ffvnBandLSgramAve}
\end{center}
\end{figure}
Figure~\ref{ffvnBandLSgramAve} shows the averaged spectrogram of 5,000 FFVN examples.
The image illustrates how the duration defined in the table shapes the averaged spread of power in each frequency band.

\section{Information hiding using kurtosis modification}\label{ss:hiding}
The level distribution of high-quality speech signals is highly non-Gaussian.
For example, the running kurtosis values have significantly higher values than those of Gaussian signal\cite{Kawahara2010intrsp}.
The following equation defines the running kurtosis $\kappa(t)$:
\begin{align}
\kappa(t) & = \frac{\mu_4(t)}{\mu_2^2(t)} \\
\mu_n(t) & = \int_{-T_w/2}^{T_w/2} w_n(\tau)s^{n}(t-\tau) d\tau \\
w_n(t) & = \frac{w^n(t)}{\int_{-T_w/2}^{T_w/2} w^n(\tau) d\tau} ,
\end{align}
where $w(t)$ represents the time window for the original signal, such as Hann window and $n$ represents the exponent, 2 or 4.

FVN and FFVN  are impulse responses of all-pass filters.
It means that they are examples of time stretched pulse (TSP).
Convolution of an FVN (or FFVN) $h_\mathrm{fvn}[n]$ and its time-reversed version $h_\mathrm{fvn}[-n]$ yields a unit pulse.
Note that we use a symbol with parentheses $x[n]$ to explicitly represents that the signal $x$ is a discrete time signal.
\begin{align}
\delta[n] & = h_\mathrm{fvn}[n] \ast h_\mathrm{fvn}[-n] ,
\end{align}
where $\delta[n]$ represents the Kronecker's delta function ($\delta[0] = 1$,  and $\delta[n] = 0$, for $n \ne 0$).
The operator $\ast$ represents convolution.
Let $x[n]$ represent a discrete time signal. Then, it follows:
\begin{align}
y[n] & = h_\mathrm{fvn}[n] \ast x[n] \\
h_\mathrm{fvn}[-n] \ast y[n] & = h_\mathrm{fvn}[-n] \ast  h_\mathrm{fvn}[n] \ast x[n] \nonumber \\
& = (h_\mathrm{fvn}[-n] \ast  h_\mathrm{fvn}[n]) \ast x[n] \nonumber \\
& = \delta[n] \ast x[n] = x[n]
\end{align}
This indicates that the time reversed FVN behaves like a key which recovers the original signal.

\begin{figure}[tbp]
\begin{center}
\includegraphics[width=0.9\hsize]{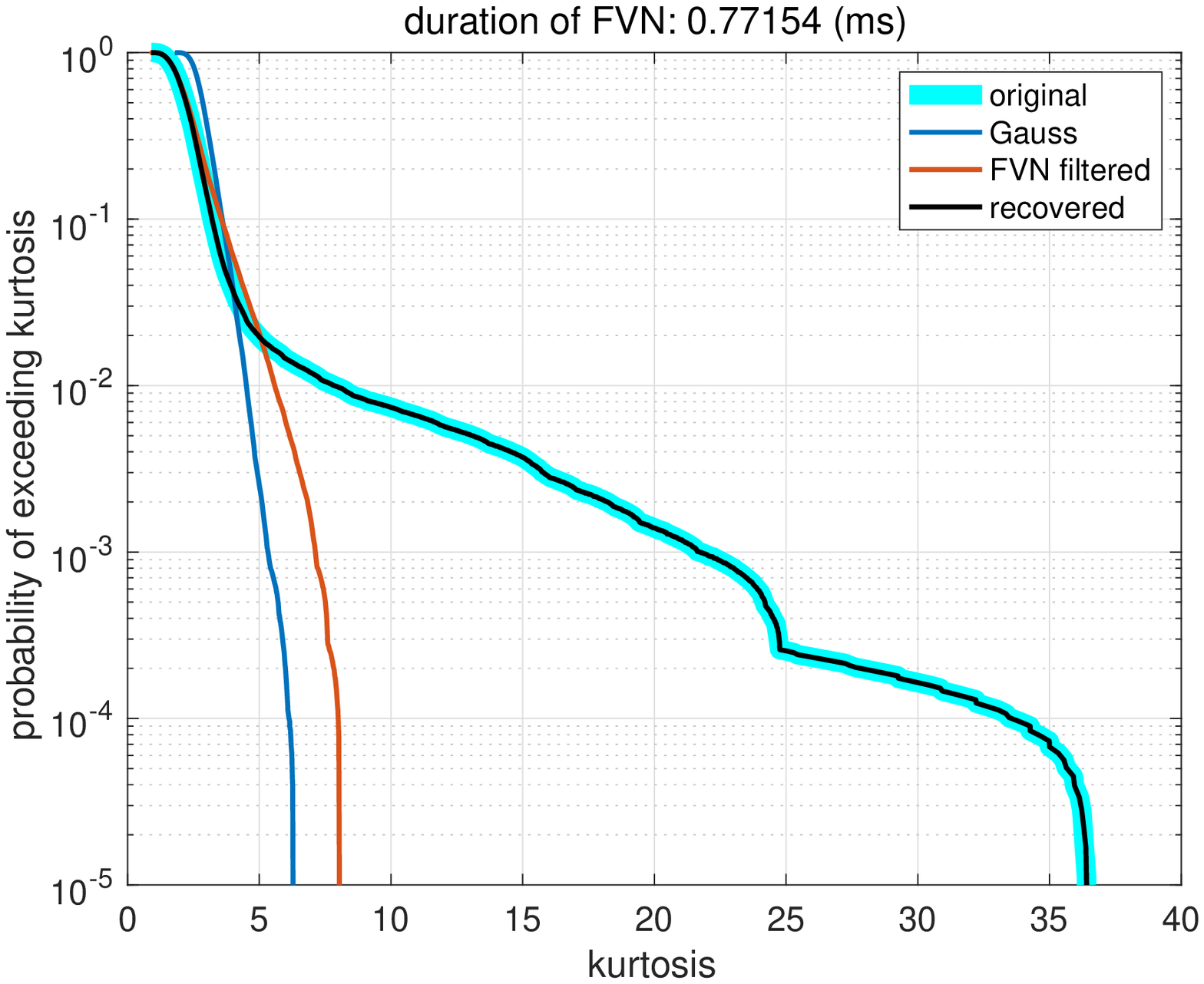}
\includegraphics[width=0.9\hsize]{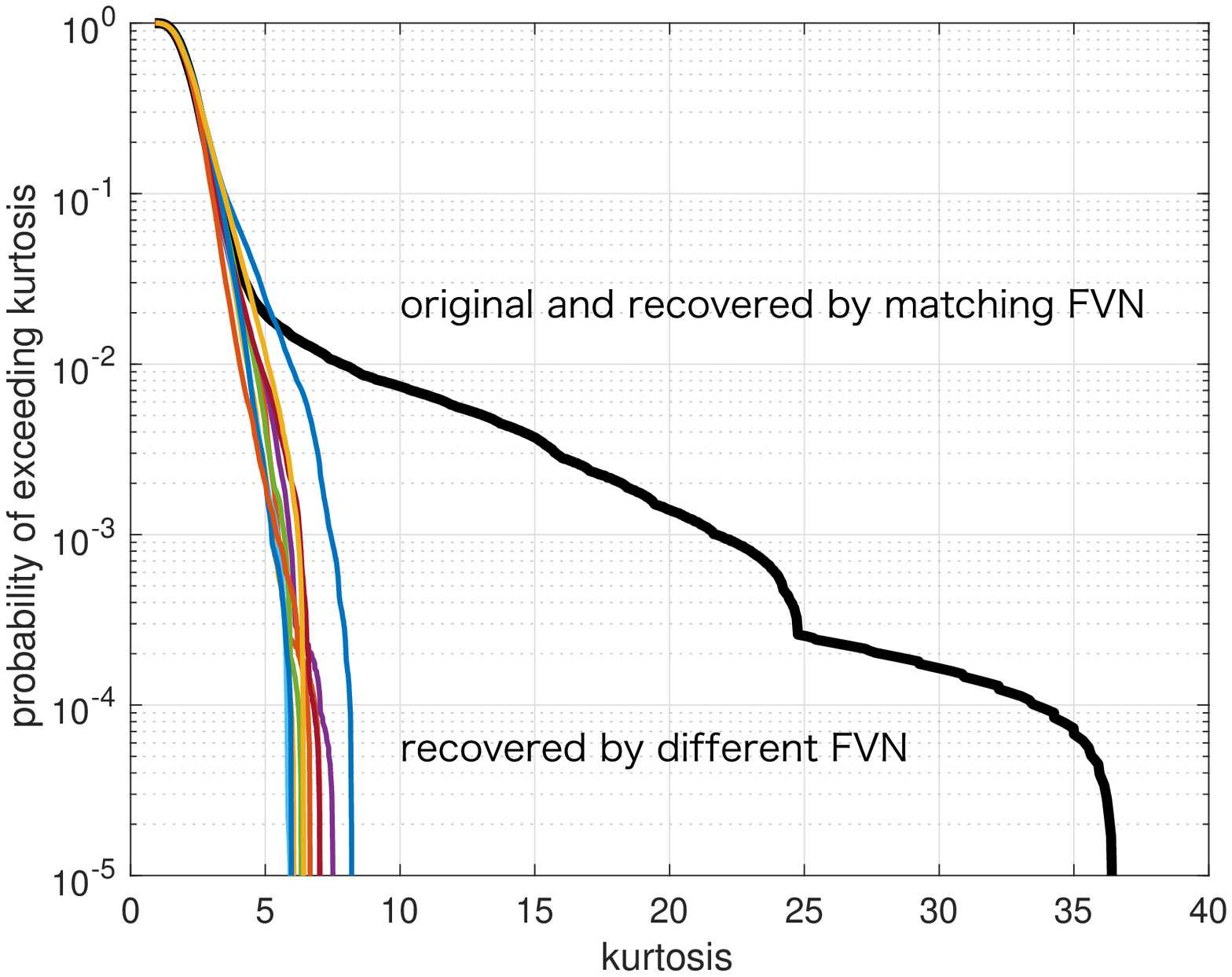}
\caption{Running kurtosis of the original speech and processed signals.
The upper plot shows the original and the processed by an FVN.
It also shows kurtosis of Gaussian and the recovered speech using the time reversed FVN.
The lower plot shows kurtosis processed by the different FVNs.}
\label{kurtosisFVNtest}
\end{center}
\end{figure}
Figure~\ref{kurtosisFVNtest} shows an example how it works.
The horizontal axis shows kurtosis. 
The original speech sample is \texttt{arctic\_a0023.wav} in the CMU ARCTIC database\cite{Kominek2004}.
The upper plot shows that
the original signal consists of about one percent of running kurtosis that is higher than 10.
It also shows that
Gaussian and FVN filtered signal rarely exceed kurtosis level 10.

The lower plot shows the original and recovered kurtosis using the matching FVN and
different FVN signals.
The figure shows kurtosis values of recovered signal using different FVNs so not exceed 10
while about one percent of kurtosis of the original signal exceeds the value 10.

These results suggest that filtering using FVN can be useful for information hiding, for example, tampering detection.
This FVN filtering itself is also useful for making instantaneous amplitude distribution of speech close to Gaussian.

\section{Excitation signal morphing}\label{ss:ifvn}
In this section, we introduce two signals from FVN (as well as FFVN).
One signal sounds like a periodic signal.
The other signal sounds like noise.
We introduce how to morph these signals seamlessly.

\subsection{Two signals: frozen IFVN and random IFVN}
Placing the same FVN or the same FFVN on a time axis repeatedly with the same interval yields a periodic sound.
We call it a frozen IFVN (Iterative Frequency domain Velvet Noise).\footnote{This naming is after IRN\cite{yost1996jasa}.}
Placing an FVN, or an FFVN generated using a different random number at each repetition yields a random signal.
We call it a random IFVN.
When the duration of each constituent FVN or FFVN is short enough regarding the repetition interval, the random IFVN
generated from different random numbers sounds like stationary white noise.

The group delay plot visualizes the difference between the frozen IFVN and random IFVN.
\begin{figure}[tbp]
\begin{center}
\includegraphics[width=0.99\hsize]{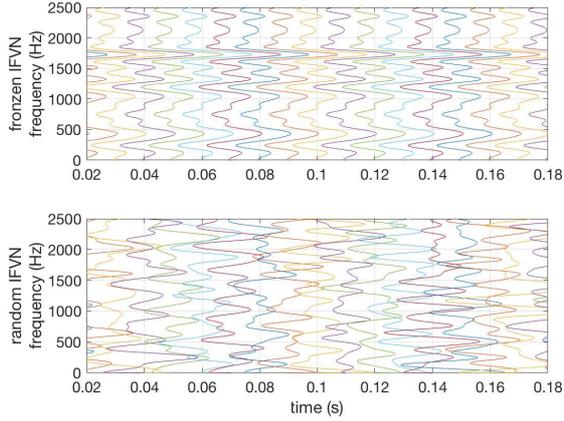}
\caption{Frozen IFVN and random IFVN examples.
Each group delay sample of IFVN is separated by 10~ms. }
\vspace{-4mm}
\label{frAndrnIFVNexmpl}
\end{center}
\end{figure}
Figure~\ref{frAndrnIFVNexmpl} is an example visualization of a frozen IFVN and a random IFVN.
The horizontal axis represents the time, and the vertical axis represents the frequency.
Because the group delay has the dimension of time, the group delay of each generated IFVN
is placed on the time axis at each location.
Note that the group delays of the random IFVN are randomly overlapping.
These overlapping yields randomized waveform.

The signals used in this simulation are generated under the following condition.
The sampling frequency is 44,100~Hz.
The FFT buffer length is 16,384.
The maximum phase deviation $\varphi_\mathrm{maz} = \pi/2$.
The support of the unit phase manipulation $B=100$~Hz.
The average frequency segment length $F_d=20$~Hz.
The fundamental frequency is 100~Hz.

\begin{figure}[tbp]
\begin{center}
\includegraphics[width=0.99\hsize]{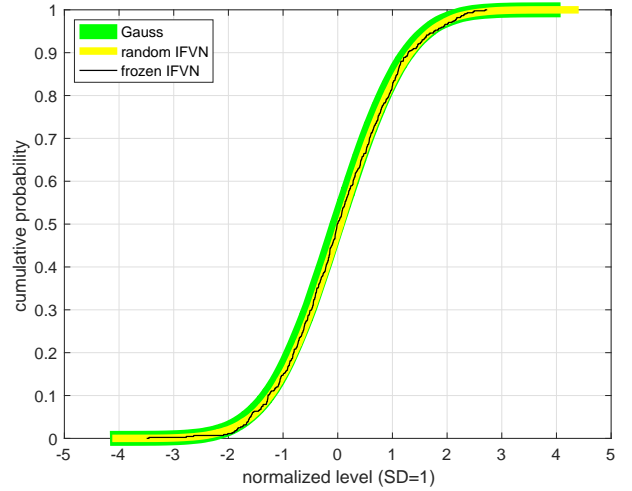}
\caption{Normalized cumulative distribution of instantaneous level of
a frozen IFVN and a random IFVN.
The Gaussian signal is also shown for reference.
}
\label{cumDistIFVN}
\end{center}
\end{figure}
Figure~\ref{cumDistIFVN} shows the cumulative probability distribution of the instantaneous level of IFVNs.
Note that the degree of freedom of frozen IFVN is smaller than that of the random IFVN.
Overlap of these plots illustrates that their distributions are Gaussian.

\subsection{Morphing by phase interpolation}
The morphing of frozen IFVN and random IFVN is multiplicative because their phases are additive in the exponent of the signal.
In other words, the phase is additive as far as it is unwrapped.
The procedure used to generate FVN or FFVN assures that their phase is not wrapped.
The following equation generates a morphed component of IFVN $x_\mathrm{mrph}(t)$.
\begin{align}
x_\mathrm{mrph}(t) & = \mathcal{F}^{-1}\left[\exp\left(j (r \theta_\mathrm{rn}(\omega) + (1-r) \theta_\mathrm{fz}(\omega)) \right) \right] ,
\end{align}
where $0 \le r \le 1$ represents the mixing coefficient of morphing.
The functions $\theta_\mathrm{rn}(\omega)$, and $\theta_\mathrm{fz}(\omega)$ represent
the phase function of an element of the random IFVN and the frozen IFVN, respectively.
The operation $ \mathcal{F}^{-1}[X]$ represents inverse Fourier transform of $X$.

\begin{figure}[tbp]
\begin{center}
\includegraphics[width=0.99\hsize]{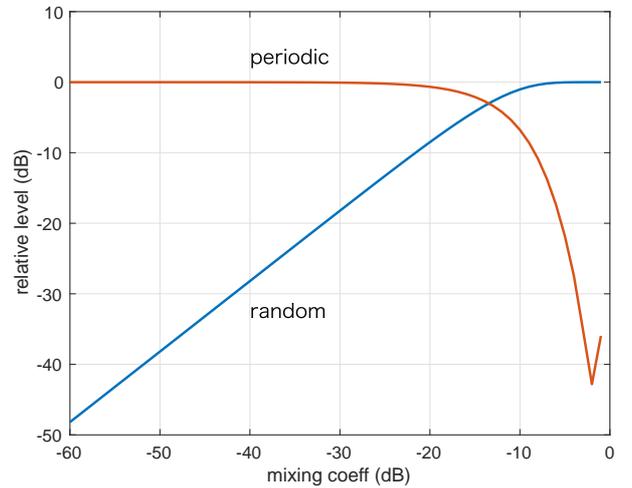}
\caption{Random and periodic component level dependencies on morphing coefficient.
The horizontal axis represents logarithmic conversion of the mixing coefficient $20\log_{10}(r)$ (dB).}
\label{morphingAndPRratio}
\end{center}
\end{figure}
Figure~\ref{morphingAndPRratio} shows how the mixing coefficient $r$ controls
the levels of the periodic and random component.
The horizontal axis represents the logarithmic conversion of the mixing coefficient. That is $20\log_{10}(r)$
(in dB).
Figure~\ref{morphingAndPRratio} is a simulation result using test signals two seconds long with 44,100~Hz sampling frequency.
This simulation result provides a method to control the periodic to random ratio directly.
\begin{align}
G_\mathrm{dBPR}(r) & = 10 \log_{10}\left[ \frac{\int_D |x_\mathrm{mrph}^{(Pr)}(t; r)|^2 dt}{\int_D |x_\mathrm{mrph}^{(Rn)}(t; r)|^2 dt}\right] \\
r(\eta_\mathrm{dBPR}) & = G^{-1}(\eta_\mathrm{dBPR}) ,
\end{align}
where $x_\mathrm{mrph}^{(Pr)}(t; r)$ and $x_\mathrm{mrph}^{(Rn)}(t; r)$ represent the periodic and random 
component in the morphed IFVN $x_\mathrm{mrph}(t; r)$, respectively.
The parameter $r$ is the mixing coefficient for morphing.
Using the inverse function $G^{-1}(\eta_\mathrm{dBPR})$ provides the appropriate mixing coefficient $r$ for the
given value of the periodic to random ratio $\eta_\mathrm{dBPR}$ in dB.

\begin{figure}[tbp]
\begin{center}
\includegraphics[width=0.99\hsize]{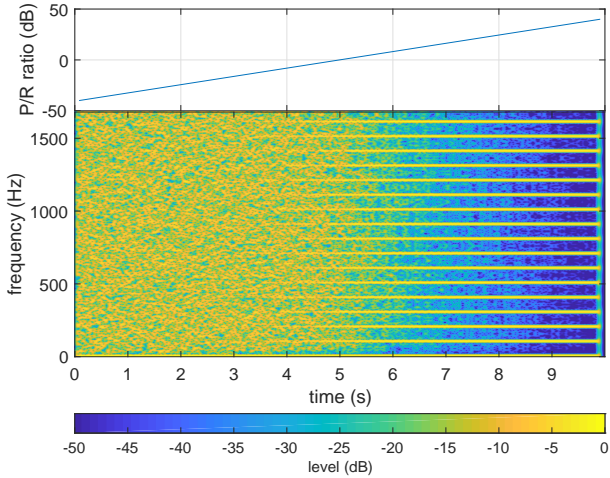}
\caption{Spectrogram of the morphed IFVN.
The top panel shows the periodic to random ratio $\eta_\mathrm{dBPR}$ in dB.}
\label{morphedIFVN10sSgram}
\end{center}
\end{figure}
Figure~\ref{morphedIFVN10sSgram} shows the spectrogram of
the morphed IFVN.
The periodic to random ratio $\eta_\mathrm{dBPR}$ starts from -40~dB and linearly increases to 40~dB.
The spectrogram indicates that the ratio control is implemented correctly.
This spectrogram uses \texttt{nuttallwin} with 200~ms length and 5~ms frameshift.

\section{Localized burst noise (FVN)}\label{ss:burst}
The location of noise burst in each glottal cycle has impact on perceived SNR (and possibly timbre),
especially for low-pitched voices\cite{Skoglund2000ieeetrans}.
An FVN with relatively short duration is appropriate for this purpose.
Based on the fact that the relative phase dependency of burst noise detection does not
show the dependency for high-pitched voices\cite{Skoglund2000ieeetrans}, the minimum duration does not need to be shorter than 0.5~ms.

\begin{figure}[tbp]
\begin{center}
\includegraphics[width=0.99\hsize]{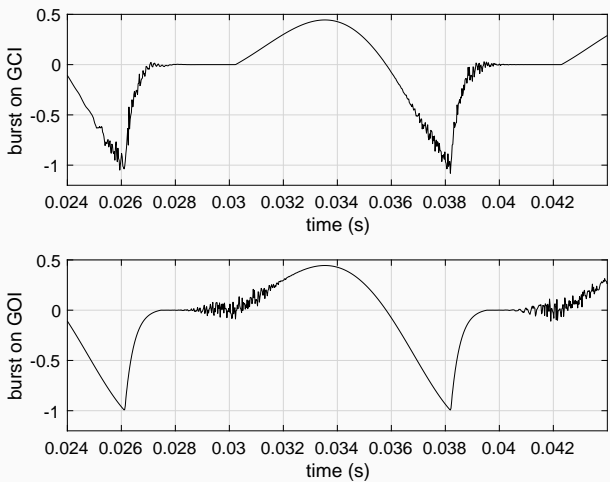}
\caption{FVN placement examples. The upper plot shows
the L-F model-based glottal wave\cite{fant1985four} with the added FVN on CGI.
The lower plot shows the waveform with FVN on GOI.
 }
\vspace{-4mm}
\label{burstOnglf}
\end{center}
\end{figure}
Figure~\ref{burstOnglf} shows the glottal waveform with the added FVN.
The antialiased L-F model generated the glottal source waveform.
The L-F model used the following parameter;
$t_p=0.4621, t_e= 0.6604, t_a=0.0270, \mbox{and,} t_c=0.7712$.
This set of parameters represents ``breathy'' voice quality\cite{Childers1991}.
The sampling frequency was 44,100~Hz.
The fundamental frequency $f_\mathrm{o}(t)$ with vibrato used the following equation:
\begin{align}
\log_2(f_\mathrm{o}(t)) & = \log_2(f_\mathrm{base}) + \frac{d_\mathrm{cent}}{1200} \sin(2 \pi f_\mathrm{vib} t) ,
\end{align}
where $f_\mathrm{vib}$ represents the vibrato frequency (5.2~Hz in this simulation)
and $d_\mathrm{cent}$ represents the depth of the vibrato (10~cent in this simulation).
The carrier frequency (center frequency of vibrato) is 82.41~Hz (musical note E2).

The FVN generation used the following parameter;
$B=2000$~Hz, $F_d=400$~Hz and $\varphi_\mathrm{max}=\pi / 2$.
The duration of the generated FVN was 0.78~ms.
The ERL of FVN was 2.7~ms.

\begin{figure}[tbp]
\begin{center}
\includegraphics[width=0.99\hsize]{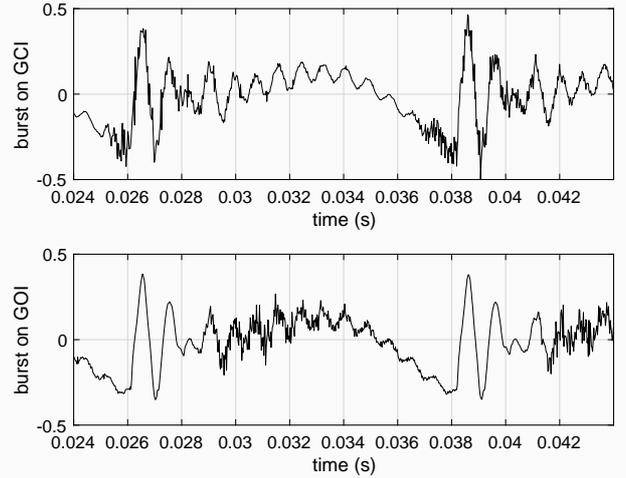}
\caption{FVN placement examples. The upper plot shows
 the generated speech waveform using the excitation with the FVN on CGI.
 The lower plot show it with the FVN on GOI.
 }
\vspace{-4mm}
\label{burstOnSp}
\end{center}
\end{figure}
Figure~\ref{burstOnSp} shows synthesized speech examples
using the speech production simulator used in SparkNG.
The impression of added FVNs are visible in these synthesized signals.

These synthesized speech samples are liked to the demonstration page.
Note that the amplitude of FVNs are 1/3 of these plots.
The averaged SNRs are the same in these two examples.
However, perceived SNR levels or at least perceived timbre of noises are different.

\end{document}